\RequirePackage{fix-cm}
\documentclass[smallextended]{svjour3}       % onecolumn (second format)
\smartqed  % flush right qed marks, e.g. at end of proof
\usepackage{graphicx}
\usepackage[super]{nth}
\usepackage{amsmath}
\usepackage{amsfonts}
\usepackage{xcolor}
\usepackage{multirow}      
\usepackage{multicol}
\usepackage{flushend}
\usepackage{natbib}
\usepackage[caption=false,font=footnotesize]{subfig}
\usepackage{url}
\usepackage{lineno}
\begin{document}

\title{Automatic Extraction of Urban Outdoor Perception from Geolocated Free-Texts}
%\subtitle{}


\author{Frances A. Santos \and
        Thiago H. Silva \and
        Antonio A. F. Loureiro \and
        Leandro A. Villas
}


\institute{Frances A. Santos \at
              University of Campinas, Brazil \\
              \email{frances.santos@ic.unicamp.br}           %  \\
           \and
           Thiago H. Silva \at
              Universidade Tecnologica Federal do Parana, Brazil\\
              \emph{Current address:} University of Toronto, Canada\\
               \email{th.silva@utoronto.ca}
           \and
           Antonio A. F. Loureiro \at
              Federal University of Minas Gerais, Brazil
           \and
           Leandro A. Villas \at
              University of Campinas, Brazil
}

\date{Received: date / Accepted: date}

\maketitle

\begin{abstract}
The automatic extraction of urban perception shared by people on location-based social networks (LBSNs) is an important multidisciplinary research goal. One of the reasons is because it facilitates the understanding of the intrinsic characteristics of urban areas in a scalable way, helping to leverage new services. However, content shared on LBSNs is diverse, encompassing several topics, such as politics, sports, culture, religion, and urban perceptions, making the task of content extraction regarding a particular topic very challenging. Considering free-text messages shared on LBSNs, we propose an automatic and generic approach to extract people's perceptions. For that, our approach explores opinions that are spatial-temporal and semantically similar. We exemplify our approach in the context of urban outdoor areas in Chicago, New York City and London. Studying those areas, we found evidence that LBSN data brings valuable information about urban regions. To analyze and validate our outcomes, we conducted a temporal analysis to measure the results' robustness over time. We show that our approach can be helpful to better understand urban areas considering different perspectives. We also conducted a comparative analysis based on a public dataset, which contains volunteers' perceptions regarding urban areas expressed in a controlled experiment. We observe that both results yield a very similar level of agreement.
\end{abstract}

\section{Introduction}
\label{intro}
Cities are not just a place composed of buildings, streets and people living together, but also a place where people have different experiences. It is known that the visual quality of urban areas and other characteristics, such as crime rate, and noise level, can evoke different perceptions about the urban area \citep{nasar1990evaluative}. Places with tourist attractions, for example, can be appreciated by tourists but avoided by residents performing their daily routines. In this direction, studies have shown that urban perception may affect people's behavior \citep{ross2001neighborhood, keizer2008spreading}. In Section \ref{sec:related}, we introduce the concept of urban perception and also how it might be important to help us understand the latent aspects of urban areas.

Capturing urban perception is not an easy task. One of the challenges is to obtain appropriate data. Field surveys and sensory walks are traditional tools to collect the opinions of people about places \citep{nasar1990evaluative, henshaw2013urban, daniele2015smelly}. Nevertheless, those strategies could be time-consuming because they usually demand a high amount of time to interview and observe volunteers and collect a considerable amount of perception samples. Thus, those strategies make it difficult to perform this type of analysis for a high number of urban areas. We explore location-based social network (LBSN) data to tackle this challenge, like other recent studies discussed in Section \ref{sec:related}. 

The number of people that use LBSNs to share their experiences, knowledge, criticism, and opinions on the most diverse topics is expressive \citep{pew2018,washpost2019}. Thus, LBSNs become an interesting choice to capture urban perception. The content is rich, enabling us to find general comments on almost everything. However, while LBSNs can help in traditional approaches' scalability issues, their use for urban perception extraction is not simple. Obtaining useful urban perceptions shared by individuals on LBSNs is challenging.

Researchers have shown evidence that LBSNs offer relevant urban perceptions  \citep{daniele2015smelly,aiello2016chatty}. Nevertheless, these previous efforts consider LBSN data to extract perceptions about some specific aspects, such as the smell of the environment \citep{daniele2015smelly} and urban sounds \citep{aiello2016chatty}. Moreover, previous studies combine traditional approaches, such as sensory walks, with LBSN data to extract the urban perceptions and, thus, lack flexibility and face scalability issues. Our work tackles those issues by offering an approach to extract urban perceptions from free-texts, without demanding traditional methods, such as sensory walks. Besides, it could be adapted to work on a different subject while still using only public LBSN content. We demonstrate our approach by considering real data from public reviews of places made on Google Places\footnote{https://cloud.google.com/maps-platform/places. Last accessed on 12 September 2020.}, and Foursquare\footnote{https://pt.foursquare.com/trip-tips. Last accessed on 12 September 2020.}, as well as public tweets from Twitter. All these data refer to Chicago and New York City, USA, and London, UK. With that, we show our approach's potential to understand the intrinsic characteristics of areas according to the people's perception. 

The main contributions of this study are:
\begin{itemize}

\item An automatic strategy to capture some of the most critical perceptions related to urban outdoor areas from geolocated text. An essential step of our approach is to construct a dictionary, namely UOP-dictionary, which organizes the main descriptive words (i.e., adjectives) used by people to qualify their experiences in urban outdoor areas of cities (Section \ref{sec:UOP-dict}).

\item An unsupervised clustering algorithm to identify content shared by individuals in free-text messages that are spatial-temporal and semantically similar. This procedure is fundamental to uncover collective perceptions of groups of people who have visited the same place in the same period (Section \ref{sec:clustering}).

\item A set of experiments to evaluate the extracted perceptions of different urban outdoor areas. We have used a Twitter dataset to demonstrate the potential of our approach to uncover the urban perceptions of outdoor spaces that emerge from LBSNs. Studying several scenarios based on LBSN data, we found valuable insights indicating the potential to better understand urban areas in many aspects. We contrasted our results with Place Pulse 2.0, a controlled experiment that contains volunteers' perceptions expressed in different urban outdoor areas \citep{dubey2016deep}. We observe that our approach yields results very similar to those indicated by Place Pulse (Section \ref{sec:results}).
\end{itemize}

\section{Contextualization and Related Work}
\label{sec:related}
Human perception is a complex process that relies on several factors, such as culture and age. There are many definitions of perception; for instance, \citet{robbins2003organizational} defines it as the process by which individuals organize and interpret their sensory impressions to give meaning to their environments. \citet{pareek1988organizational} explains the same concept as the process of receiving, selecting, organizing, checking, and reacting to sensory stimuli or data. 

While there are several definitions, we agree that a key aspect of perception is the process of interpreting sensory stimulus, coming from the five senses: sight, taste, smell, touch, and sound. Thus, the sensed information is sent to our brain. Then the perception is how we interpret the sensed information (or sensations) to make sense of everything in our surroundings. In this way, the perception is related to sensation but is much more than that.

Each individual can perceive the same place, object or situation differently based on relevant aspects to him/her. The main reason for this is that people interpret things by considering their past experiences and the values they consider appropriate. The perceptions of a group can also influence one's perception. Thus, perceptions can differ from person to person.

We can obtain individuals' perceptions of virtually anything. However, in this study, aligned with others, we focus on the individuals' perception of urban environments, called \emph{urban perception}. More specifically, we are interested in perceptions regarding urban outdoor areas. Those types of areas, such as parks, streets, and plazas, may offer people the opportunity of having diverse experiences, and, for this reason, could trigger different perceptions among their visitors. 

A traditional approach to capture urban perceptions is sensory walks, where volunteers answer surveys that contribute to building the common sense about inherent characteristics of cities, which is a common approach to collect features regarding urban areas \citep{keizer2008spreading, henshaw2013urban}. This approach provides fine-grained data about urban spaces' perceptions, but it does not scale easily \citep{daniele2015smelly, dubey2016deep}. To overcome this problem, a possible strategy is to extract urban characteristics shared by people in online sources, such as crowdsourcing systems and location-based social networks (LBSNs), which are useful for understanding and tackling different problems.

Using LSBN data, we can study a variety of urban outdoor perceptions, ranging from perceived safety and traffic conditions to aesthetic characteristics and the cost of living in urban areas. Some studies concentrate their efforts to understand the factors present in the environment responsible for triggering such perceptions.

Towards that direction, \citet{quercia2014aesthetic} collected people's perceptions about photos taken in the streets of London. They found a correlation among colors, texture and visual patches of those photos with beautiful, happiness, and quiet perceptions. In the same direction, \citet{naik2014streetscore} proposed an approach for predicting the perceived safety of cities exploring images from different sites. Similarly, \citet{porzi2015predicting} and \citet{de2016safer} demonstrated that computer vision techniques could also identify human perception and predict judgments of safety from images of urban scenes. \citet{dubey2016deep} explored a global dataset of urban images to rank street-level images of city aesthetics by also using computer vision techniques. Their results showed that urban perception data on a worldwide scale could be extracted from online photos. 

\citet{daniele2015smelly} explored sensory walks to collect the citizens' perception concerning the smell of the environment. This offline process has enabled the creation of a dictionary with urban-smell related words, which is used to discover messages related to odor perceptions into social media data. \citet{quercia2016emotional} presented a follow-up of this study, where the authors investigated, among other things, the relationship between the predominant color of the image (visual perception) and the smell associated with the image (olfactory perception). Also, considering the citizens' perception regarding urban smell, \citet{hsu2019smell} developed an online crowdsourcing system to enable Pittsburgh's community to report odors and uncover where those odors are frequently concentrated.

\citet{aiello2016chatty} explored the influence that urban sounds have on the way people perceive places. The authors created a dictionary with sound-related words to discover the urban sounds responsible for triggering the people's perceptions and emotions, which was used to mine LBSN data related to urban sounds. In this way, London and Barcelona's streets could be mapped with one of the six considered categories (transport, mechanical, nature, human, music, indoor) of the dictionary created.

In the literature, different modeling strategies have been used to extract information about a given subject. Several proposals have characterized the urban spaces according to socio-cultural activities \citep{steiger2016exploration}, functional zones \citep{yuan2015discovering, jenkins2016crowdsourcing, yao2017sensing}, semantics and sentiments (or emotions) \citep{hu2019semantic}, or points of interest \citep{jiang2016personalized}.

Recent studies have made efforts to automatically uncover and predict the perception reflected by urban areas, using, to this end, pictures of outdoor places collected from online sources. \citet{santani2018looking} collected, using mobile crowdsourcing, a total of 7,000 geotagged images from three Mexican cities, which were used to gather opinions from 144,000 ``workers'' of the Amazon Mechanical Turk. With this, the authors proposed a deep learning framework to learn and characterize the urban perception based on visual features present in the observed images and the perceptions related to them. Similarly, \citet{redi2018spirit} collected geolocated Flickr pictures, and the tags attached to them, posted by individuals in Greater London. Next, the authors proposed a methodology to map London's neighborhoods' ambiance, based on a psychological taxonomy.

As some of the related studies, we also take advantage of LBSN data to facilitate the process of understanding people's perception of urban areas. We acknowledge that using only social media as a data source can be problematic in some cases, such as in smaller cities and less tourism-related places, where the extraction of relevant perception, if they are identified, can fall only within some specific topics, perhaps less informative. However, despite considering only social media data in our work, due to several reasons (e.g., easy access, public availability, large-scale data for larger cities, and rich content), the proposed approach does not depend on any exclusive social media characteristics to properly work, as other literature studies that use, for instance, hashtags, emoticons, and tags. In this sense, we consider that other kinds of data sources, such as IoT devices, smart vehicles, and government data, could be used by our approach. They can have natural language texts, geospatial, and temporal information, potentially contributing to deal with bias present in social media data.

 Moreover, our work significantly differs from previous studies in several aspects. For instance, the format in which urban perceptions can be shared on LBSN is diverse, such as via text, image, video and audio. We focus our work only on perceptions shared via text message. The proposed approach is also one of the very few ones that explore free-texts shared by users on LBSNs to extract several perceptions related to urban outdoor areas. Our approach does not require time-consuming field surveys or manual steps to extract urban outdoor perceptions.
 
 Finally, our approach also differs from social media specialized in rating places, such as Yelp, TripAdvisor, Foursquare and Google Places. These systems enable their users to share reviews about places, containing rich information regarding our area of study. However, data on such systems are usually more available to indoor places, such as restaurants, and points of interest, such as museums, rather than the more general urban areas, such as avenues, streets and squares. Thus, opinions unlinked to an establishment, i.e., perceptions shared while transiting in the city, may not be properly captured. Besides, they are heavily dependent on users' engagement who must actively participate in these systems; thus, for some unpopular places, there might be only a few, if any, data. Exploring only these sources might offer a partial view of the whole urban perception spectrum. In this way, our approach can be helpful to better understand urban areas considering different perspectives, in an automatic and generic way.

\section{Dictionary Creation}
\label{sec:UOP-dict}
The use of tweets to extract urban perception is difficult given their content diversity, which is not restricted to the subject under investigation. To overcome this problem is essential first to explore a less noisy source and, thus, learn properly the vocabulary used by people regarding outdoor areas.

In this sense, we selected outdoor urban areas in Chicago and New York City (NYC), United States, and London, United Kingdom, to collect public reviews shared before February 2017 on Google Places and Foursquare (Tips). Such systems enable us to specify the place categories\footnote{Foursquare Venue Category Hierarchy: \url{https://developer.foursquare.com/docs/build-with-foursquare/categories/}. Last accessed on 12 September 2020.}\footnote{Google Places Category: \url{https://developers.google.com/places/web-service/supported\_types}. Last accessed on 12 September 2020.}, which allow selecting just reviews about outdoor places. Additionally, we filtered out reviews not written in English. We call this dataset of \textit{Places Review}. Table \ref{tab:non-geotag-dataset} summarizes the number of collected reviews. As we can see, due to the restriction of Google Places API, the documents from Foursquare represents most of the \textit{Places Review} dataset, which has a total of 39,348 reviews. Using this dataset, we want to uncover frequently used words to qualify experiences in urban outdoor areas. This step is essential to build our urban outdoor perception dictionary, namely the UOP-dictionary. 

\begin{table}[!htp]
  \centering
  \caption{Reviews in the \textit{Places Review} dataset.}
  \label{tab:non-geotag-dataset}
  \begin{tabular}{cccc}
    \cline{2-4}
    \multicolumn{1}{l}{} & \multicolumn{1}{l}{\textbf{Chicago}} & \textbf{London} & \multicolumn{1}{l}{\textbf{NYC}} \\ \hline
    \multicolumn{1}{c|}{Foursquare Tips} & 5,085 & 7,261 & 24,921 \\
    \multicolumn{1}{c|}{Google Places} & 666 & 662 & 753 \\ 
  \end{tabular}
\end{table}

More formally, the \emph{Places Review} dataset can be defined as follows:

\begin{definition}
  A collection of documents $\mathcal{D}_{R}$, where each $doc \in \mathcal{D}_{R}$ is a document determined by a tuple $doc=(id,s,\tau)$, where $id$ is a unique identifier, $s$ is a list of sentences that comprise all content written by the user, and $\tau \in \mathbb{R}$ is a timestamp.
\end{definition}

In this work, the term sentence is used to refer to the preprocessed free-text, where numbers, Uniform Resource Locators (URLs), special characters, punctuation, and stop words were removed, and frequent English contractions, e.g., \textit{aren't} and \textit{I 'm}, were replaced by the expanded form, e.g., \textit{are not} and \textit{I am}. Next, we perform the stemming process to obtain the words in the common root form\footnote{https://www.nltk.org/howto/stem.html. Last accessed on 12 September 2020.}, resulting in a vector of single words' stem. Having preprocessed the documents of Place Reviews, we follow an automatic approach to create the UOP-dictionary.

First, we perform part-of-speech tagging in each sentence $s \in doc, \forall doc \in \mathcal{D}_{R}$, to classify which tag, among noun, adjective, verb, pronoun, etc., is most likely for words of $s$. After labeling all words, we extract a set of words considered qualifiers, i.e., marked with the adjective tag, performing a double-check using  WordNet\footnote{http://www.nltk.org/howto/wordnet.html. Last accessed on 12 September 2020.}, which contains a corpus of words already tagged. After that, we obtained a list with $271$ commonly used words by people to qualify outdoor places. The next step is to organize these words into categories, according to their syntactic and semantic similarity, and remove the words that do not fit well in any category, being considered outliers.

To this end, we calculate a similarity score ($score_{sim}$) for each pair of words, using for that the Word2Vec model \citep{mikolov2013distributed} and the sentiment polarity between words. Word2Vec is a model based on a neural network used to transform words into high-dimensional vectors that contain many linguistic regularities and patterns \citep{mikolov2013distributed}. It takes as input a corpus $C$ (i.e., all sentences in documents of $\mathcal{D}_{R}$), a window size $ws$, a minimum count of occurrence of the word $minCount$, and a hyper-parameter $m$ that represents the number of features. Next, the model creates a vocabulary of $n$ unique words denoted by $W$ from $C$, where each word $w \in W$ must occur at least $minCount$ times in $C$. The Word2Vec model computes for each pair of words $w_1, w_2 \in W: w_1 \neq w_2$, the probability to find them ``nearby'' into sentences of $C$. Two words are considered nearby if they are at most $ws-1$ positions between them.

Using the \textit{Places Review} dataset, where the size of \textit{corpus} is $C = 54,612$, we trained our Word2Vec model to produce the word vectors, enabling us to identify if two different words have similar contexts, checking the similarity of their word vectors (value denoted by $w2v_{sim}$). We empirically defined $ws = 8$, $minCount = 20$, and $m = 300$. The skip-gram model and hierarchical softmax worked better in our experiments; therefore, we employed them.

In addition to Word2Vec, we also compute the product between the sentiment polarity for each pair of words (value denoted by $sent_{sim}$), using NLTK VADER\footnote{https://www.nltk.org/howto/sentiment.html. Last accessed 12 September 2020.} to this end. For instance, given two words, $w_1$ and $w_2$, if both have the same sentiment polarity (negative or positive), then the $sent_{sim}$ is positive. Otherwise, the product between their sentiment polarity will be negative. Also, when one of the words has a neutral polarity, the $sent_{sim}$ becomes zero. Thus, $score_{sim}$ captures the similarity of words as follows:
\begin{equation}\label{eq:ss}
    score_{sim} = \alpha \times w2v_{sim} + (1-\alpha) \times sent_{sim}
\end{equation}
\noindent where $\alpha = 0.8$, since the $w2v_{sim}$ brings valuable information about the contextual relationship between words, whereas the $sent_{sim}$ works as a filter, either smoothing or increasing the $score_{sim}$ according to sentiment polarity of words. 

\begin{figure}[!htp]
    \centering
    \includegraphics[scale=0.25]{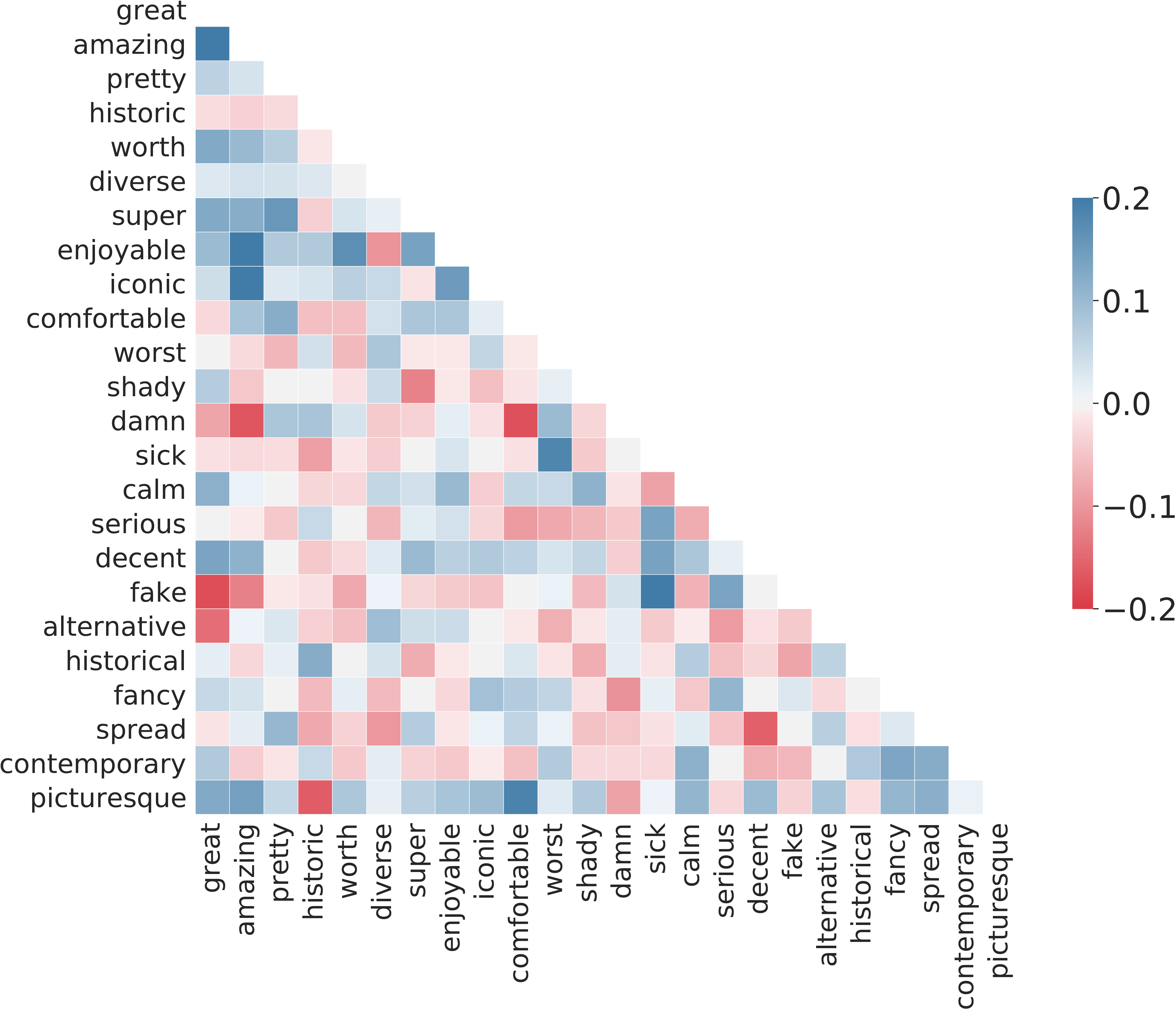}
    \caption{Correlation matrix for some randomly selected words under consideration. [Best in color]}
    \label{fig:correlation-matrix}
\end{figure}

Figure \ref{fig:correlation-matrix} shows the $score_{sim}$ for a randomly selected sample of words from the model. As we can see, pairs of similar words, such as \emph{amazing} and \emph{great}, \emph{iconic} and \emph{enjoyable}, and \emph{damn} and \emph{worst}, have high $score_{sim}$ value (dark blue). On the other hand, pairs of words with low $score_{sim}$ value (dark red), such as \emph{great} and \emph{fake}, and \emph{damn} and \emph{amazing}, represent dissimilar words.

In this way, we can build a graph $G = (V, E, \omega)$, where $V$ is the set of vertices that represent the words of the model, $E$ is the set of connections (edges) between the vertices, where each $e \in E$ has a real weight $\omega_e \in \omega$, which are defined by Eq.\ \eqref{eq:ss}. In order to avoid meaningless connections between pairs of vertices, we calculate a threshold to eliminate ``light weight'' edges as follows. For each vertex $u \in V$, being $E_u$ a set of edges incident to vertex $u$, i.e.,  $E_u = \{(u,v): v \in V, v \neq u\}$, the threshold $thresh_u$ is defined by Eq.\ \eqref{eq:thresh}:
\begin{equation}\label{eq:thresh}
    thresh_u = \overline{X_u} + \beta \times \sigma_u
\end{equation}
where $\overline{X_u}$ denotes the weights mean of edges incident to $u$:
\begin{equation}\label{eq:mean}
    \overline{X_u} = \frac{\sum \omega_{u,v}}{\arrowvert V \arrowvert -1}
\end{equation}
where $\sigma_u$ denotes the standard deviation: 
\begin{equation}\label{eq:std}
    \sigma_u = \sqrt{\frac{\sum (\omega_{u,v}-\overline{X_u})^2}{\arrowvert V \arrowvert -2}}
\end{equation}
and $\beta \geq 0$ defines the level of restriction. Thus, if $\beta=0$ then $thresh_u = \overline{X_u}$, and, consequently, unmeaningful connections may remain on the graph. To avoid that, we consider a conservative value to $\beta$ ($\beta = 1.13$), in order to maintain only meaningful edges. After removing from $E$ the edges with the weight less than or equal to $thresh_u, \forall u \in V$, we obtain a graph $G'=(V', E', w')$, where $V' \subset V$ ($\arrowvert V' \arrowvert = 261$, after removing ten isolated vertices), $E' \subset E$ ($\arrowvert E' \arrowvert = 3,485$), and $w' = \{w_e: w_e \in w, \forall e \in E'\}$. With graph $G'$, we are able to group the vertices according to their similarity, in terms of connections.

To achieve this, we identify overlapping communities in the graph $G' $ based on the clique percolation method \citep{palla2005uncovering}. Thus, a $k$-clique community, or simply community, is the union of all cliques of size $k$ that can be reached through adjacent (sharing $k-1$ vertices) $k$-cliques \citep{palla2005uncovering}. The clique percolation method is particularly interesting because vertices weakly connected are not placed in any community, thus removing possible noises. Also, because it allows the existence of overlapped vertices, i.e., present in multiple communities, we avoid splitting large communities into small ones with similar contexts. We empirically defined $k = 6$, which resulted in eight communities of cohesive vertices (words). These communities comprise our dictionary called UOP-dictionary.

\begin{figure}[!tp]
    \centering
    \includegraphics[scale=0.35]{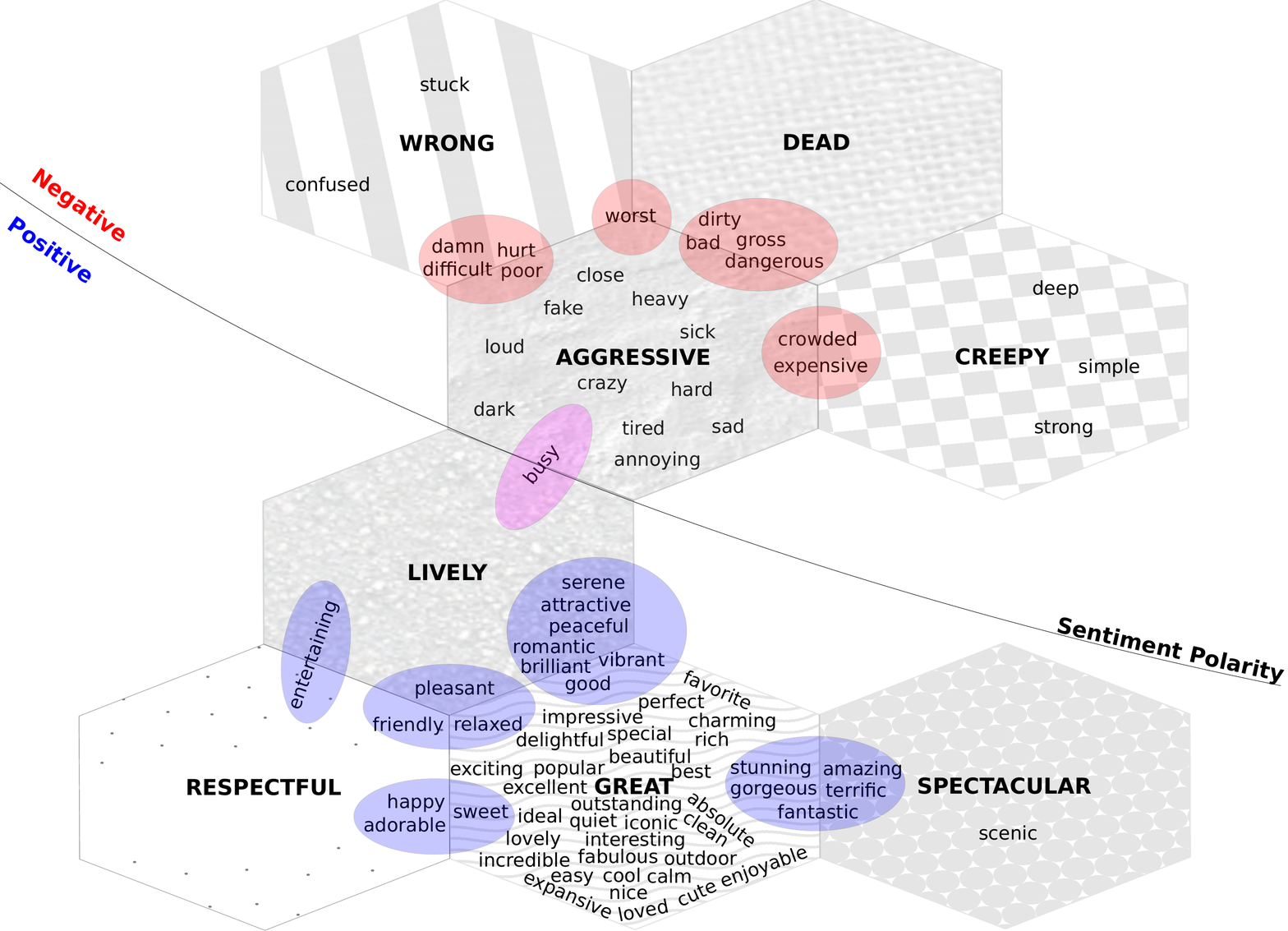}
    \caption{Illustration of the UOP-dictionary.}
    \label{uop-dict}
\end{figure}

An illustration of this dictionary is shown in Figure \ref{uop-dict}. To organize this image, we separated the communities according to their sentiment polarity. Half of the communities have words with positive polarity, and the other half has negative polarity.

Next, words of each community were placed into hexagons, where the overlapped words are highlighted with blue (positive), red (negative), or pink (transition between positive and negative) circles. To simplify the community labeling, we chose one of the available words that better represent the group's semantics --overlapped words were not selected. This dictionary is an important step to discover the perception of outdoor places based on noisy LBSN data.

\section{Extraction of Urban Perceptions}
\label{sec:clustering}
To extract the people's perception from messages shared via LBSN about urban outdoor environments, we explored public messages (tweets) of Twitter. Twitter is an online microblogging service, where people can, among other things, share short messages of a maximum size of 280 characters. Using Twitter API, it is possible to gather tweets for areas of interest delimited by bounding boxes, where a fraction of them are geotagged (the ones we consider). Geotagged tweets have been used in many applications to predict or detect events in near real time \citep{kalampokis2013understanding} and as a proxy of mobility \citep{lenormand2014cross}.

To build the LBSN dataset, we collected tweets, from January to August 2018, for Chicago, NYC, and London. We also represent the LBSN dataset as a document collection defined as:

\begin{definition}
  Collection $\mathcal{D}_{L}$, where each $doc \in \mathcal{D}_{L}$ is a document determined by a tuple $doc=(id,s,\tau,g)$, where $id$ is a unique identifier, $s$ is a list of sentences that comprise all content written by the user, $\tau \in \mathbb{R}$ is a timestamp, and $g \in \mathbb{R} \times \mathbb{R}$ is the geographic coordinates, expressed by latitude and longitude, respectively.
\end{definition}

When applying the UOP-dictionary in $\mathcal{D}_{L}$, several documents not related to individual perception about urban outdoor areas tend to be retrieved. The reason for that is because the qualifiers that compose the dictionary are not restricted to describe places. Some of them can also be used to describe other entities, such as people and things. To address this problem, we group documents that have both spatial-temporal and semantic similarities, and, thus, disregard documents non-related to urban areas, or individual perceptions unrelated to urban outdoor areas. To this end, we propose an unsupervised clustering algorithm to group documents with \emph{spatial}, \emph{temporal}, and \emph{semantic} similarities present in the data. 

First, we remove any spatial noise from the dataset containing LBSN data, called $\mathcal{D}_L$, i.e., all documents whose geolocation is coincident to others according to a threshold, $thresh_{spatial}$, are filtered out from $\mathcal{D}_L$. The $thresh_{spatial}$ indicates the acceptable maximum number of documents with a coinciding GPS location. The probability of this situation to happen in practice is low for real data, being more common in automatic processes, such as those used by web robots, as discussed in \citep{tasse2017state}. Thus, this step is essential to avoid considering invalid data. Based on the documents $\mathcal{D}_L$, we have determined a threshold, $thresh_{spatial} = 10$, to perform this filtering since most geolocations (about 90\%) have less than $10$ documents associated with them. This process produces the collection $\mathcal{D}_ L' $.

Next, we use the UOP-dictionary to label the appropriate category of perception of documents in $\mathcal{D}_ L' $.

In this way, a given document with at least one word corresponding to the UOP-dictionary can be labeled with one of the eight labels of the dictionary. If two or more words are present in the dictionary, multiple labels can be assigned to the same document. This helps to reflect the heterogeneity of perceptions, where the same area may have distinct perceptions according to the people's opinion. Alternatively, the document may not have words matching with the dictionary, which implies that it may not be related to an outdoor area qualification and, therefore, is disregarded. As a result, we have a new collection of documents $\mathcal{D}_L" $, where the documents are properly labeled with at least one community label of the UOP-dictionary.

After that, we perform a semantic similarity task. For each document $doc \in \mathcal{D}_L" $, we explore the Word2Vec model to calculate the likelihood of a $doc$ be a member of a specific community of the UOP-dictionary \citep{taddy2015document}. This procedure results in a score, ranging from $0$, very unrelated, to $100$, very similar, and we explore it to decide if the $doc$ has enough semantic similarity with the model. Based on the dataset $\mathcal{D}_L''$, we have determined a threshold, $thresh_{semantic} = 18$, to perform this classification, where most documents (about 75\%) have a score higher than $18$. To define this value, we studied the distribution of scores. This enabled us to observe a clear change on the curve when the score is $18$, indicating that after that point, documents tend to be more related to the context of interest. In fact, when evaluating documents with score values less than $18$, we start to find documents unrelated to urban outdoor areas. On the other hand, for scores slightly above $18$, we practically find only documents related to the urban outdoor context.

\begin{table*}[!tp]
\scriptsize
\centering
\caption{Number of documents after main steps of our approach.} %Algorithm \ref{alg:cluster}
\label{tab:geotag-dataset}
\begin{tabular}{cccc}
\cline{2-4}
\multicolumn{1}{l}{} & \multicolumn{1}{l}{\textbf{Chicago}} & \textbf{London} & \multicolumn{1}{l}{\textbf{NYC}} \\ \hline
\multicolumn{1}{c|}{Total of documents ($\mathcal{D}_L$)} & 803,689 & 1,364,922 & 2,690,327 \\
\multicolumn{1}{c|}{\# documents after applying the spatial filter ($\mathcal{D}_L'$)} & 181,836 & 345,100 & 620,350 \\
\multicolumn{1}{c|}{\# documents match with the UOP-dictionary ($\mathcal{D}_L''$)} &  24,951 & 58,482 & 73,062 \\
\multicolumn{1}{c|}{\# documents after applying the semantic filter ($\mathcal{D}_L'''$)} & \textbf{18,293} & \textbf{39,760} & \textbf{54,539} \\ \hline
\end{tabular}
\end{table*}

Table \ref{tab:geotag-dataset} summarizes the number of documents remaining after each step of our approach. The last line highlights the total of documents in $\mathcal{D}_L'''$, representing semantically similar documents, which are clustered according to their spatial-temporal similarity. To this end, we split the documents monthly, according to the time they were shared. Also, we used the Hierarchical Density-Based Spatial Clustering of Application with Noise (HDBSCAN) \citep{campello2013density}, for clustering the documents based on their spatial similarity. HDBSCAN extends the DBSCAN \citep{ester1996density}, using a technique to find a clustering that gives the best stability over an $\epsilon$ value \citep{campello2013density}. To explore HDBSCAN for this task is interesting since it is based on a neighborhood density to identify spatial clusters. Also, it can create clusters with different formats and sizes, and, typically, achieves satisfactory results even in the presence of noises. This clustering process results in a set of clusters $\mathcal{C}$.

\begin{figure*}[!tp]
   \centering
    \subfloat[Word cloud of the documents matched to UOP-dictionary ($\mathcal{D}_L''$).]{
        \includegraphics[scale=0.5]{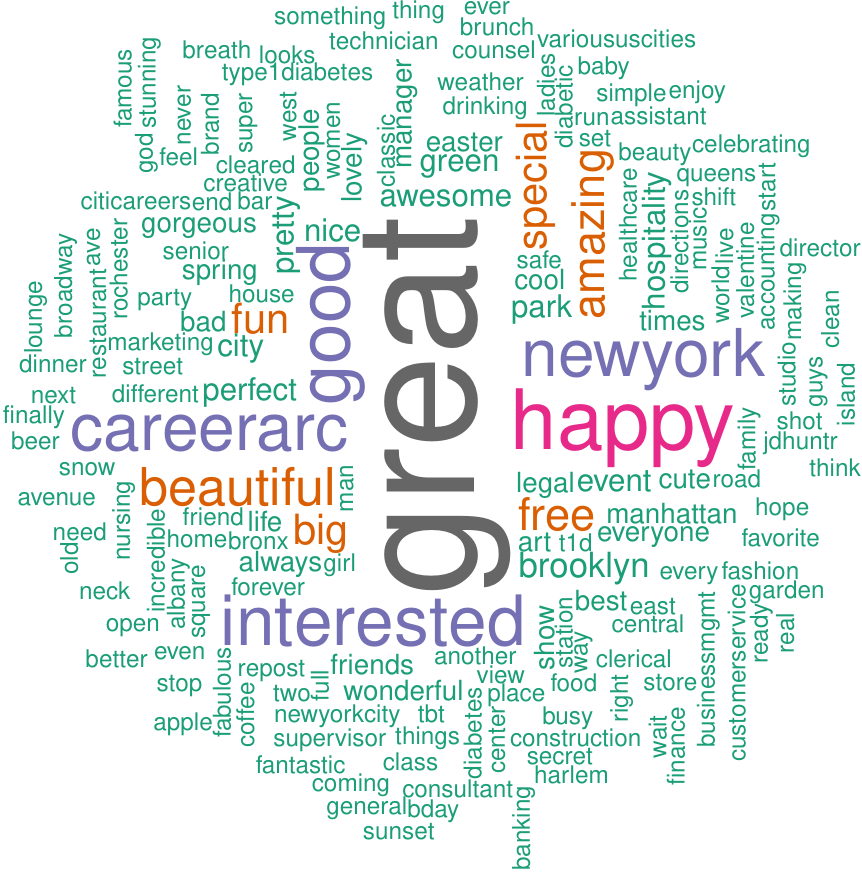}}
\hspace{0.2cm}        
    \subfloat[ Word cloud of the documents in $\mathcal{C}$.]{
        \includegraphics[scale=0.5]{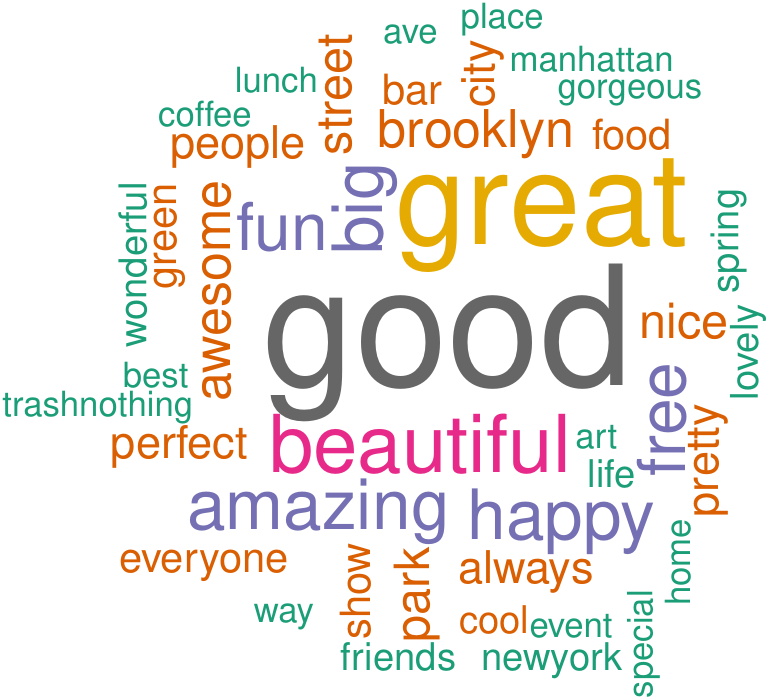}}
    \caption{Word clouds before and after applying our approach.}
    \label{fig:wordclouds}
\end{figure*}

To illustrate the potential of our approach to extract urban perceptions, Figure \ref{fig:wordclouds} shows two word clouds, where more centralized words and with larger font sizes are the most frequent ones, considering data collected from NYC. In Figure \ref{fig:wordclouds}(a), before applying our approach, we can observe that several frequent words are not related to urban outdoor areas, e.g., ``careerarc'' and ``interested''. On the other hand, the remaining documents, i.e., the documents in $\mathcal{C}$, are strongly related to urban outdoor perceptions as displayed in Figure \ref{fig:wordclouds}(b), being more suitable to be used on perception mapping.

\section{Experiments and Evaluation}
\label{sec:results}
This section discusses the results and evaluation of our proposed approach, considering Chicago, New York City (NYC) and Greater London as our scenarios (see Appendix \ref{apped_a}). For each studied city, we selected three neighborhoods with the objective of evaluating diverse areas in each city. To this end,  we selected representatives of rich/central areas, problematic areas (e.g., poor and/or violent), and residential and non-problematic areas, favoring neighborhoods with more data. 

\subsection{General Map at City Scale}
\label{sec:general_map}
\begin{figure}[!htp]
   \centering
    \includegraphics[scale=0.35]{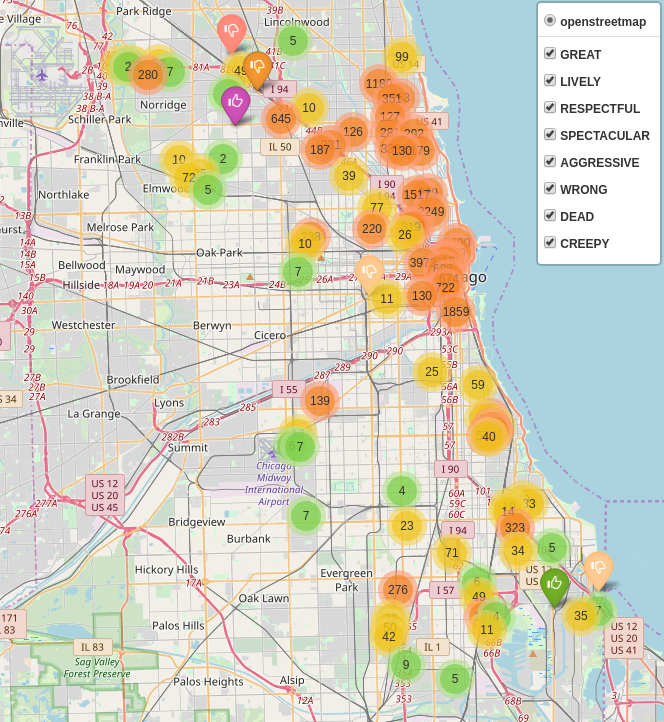}
    \caption{Screenshot of an interactive urban perception map in Chicago. This tool was built for this study to help to evaluate the perceptions found.}
    \label{fig:perception-maps-chicago}
\end{figure}

With the help of interactive maps built for this study, urban perceptions are mapped into the cities according to the number of observed collective perceptions. This tool eases the process of studying the perceptions found. For instance, Figure \ref{fig:perception-maps-chicago} shows the heat maps of clusters for each perception category separately, considering all data on the evaluated areas from January to August 2018 (see Appendix \ref{apped_b} for other cities). As we can see, there is an interactive selector to filter the desired perceptions, which help to analyze each perception individually. Besides, our application visually combines close clusters into one single representation (circle) and simplifies the display of clusters on the map. The number on a circle indicates how many points in all grouped clusters it contains -- colors from green (low) to red (high) indicate the number of points. As the zoom increases, the clusters are split, and individual markers will, eventually, appear. Such maps are useful to show the perception concentration at the city scale. 

% \begin{figure}[!tp]
%   \centering
%     \subfloat[Chicago.]{
%         \includegraphics[scale=0.22]{chicago_perception_maps.png}} 
%     \subfloat[NYC.]{
%     \hspace{0.2cm}
%         \includegraphics[scale=0.234]{ny_perception_maps.png}} \protect \\
%     \subfloat[London.]{
%         \includegraphics[scale=0.3]{london_perception_maps.png}}
%     \caption{Urban perception maps for the evaluated scenarios.}
%     \label{fig:perception-maps}
% \end{figure}

\subsection{Spatial Analysis at Neighborhood Scale}
\label{subsec:spatial_analysis}

\begin{figure}[!ht]
   \centering
   \includegraphics[scale=0.45]{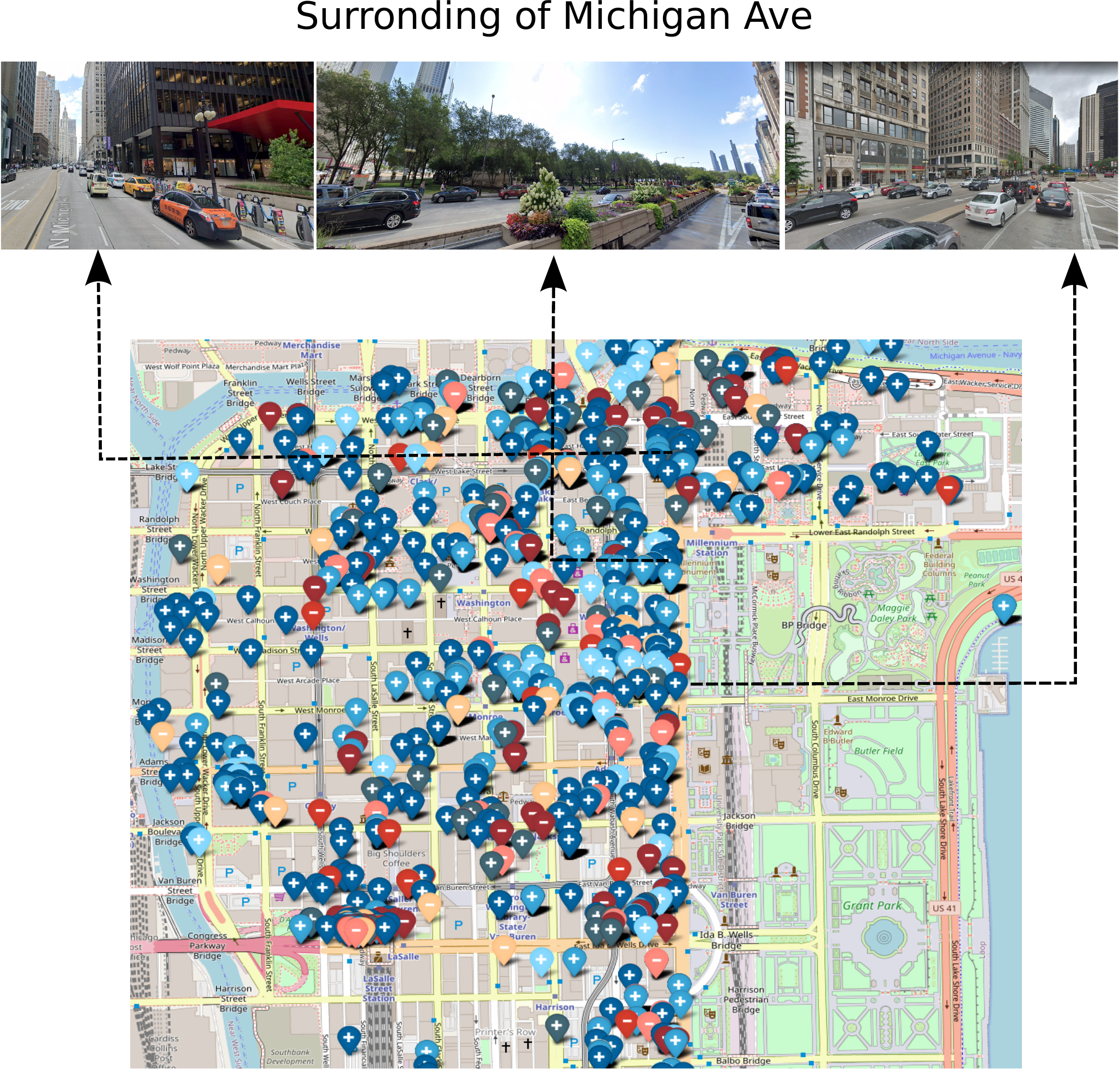}
   \caption{Perceptions map in Loop (Downtown). [Best in color]}
   \label{fig:chicago-loop-perception-maps}
\end{figure}
%Colors represent perceptions classes: dark blue (GREAT); gray (LIVELY); blue (RESPECTFUL); light blue (SPECTACULAR); dark red (AGGRESSIVE); red (WRONG); light red (DEAD); and beige (CREEPY). Symbols ``Plus'' or ``Minus'' represent positive and negative perceptions, respectively.
\begin{figure}[!hb]
   \centering
   \includegraphics[scale=0.35]{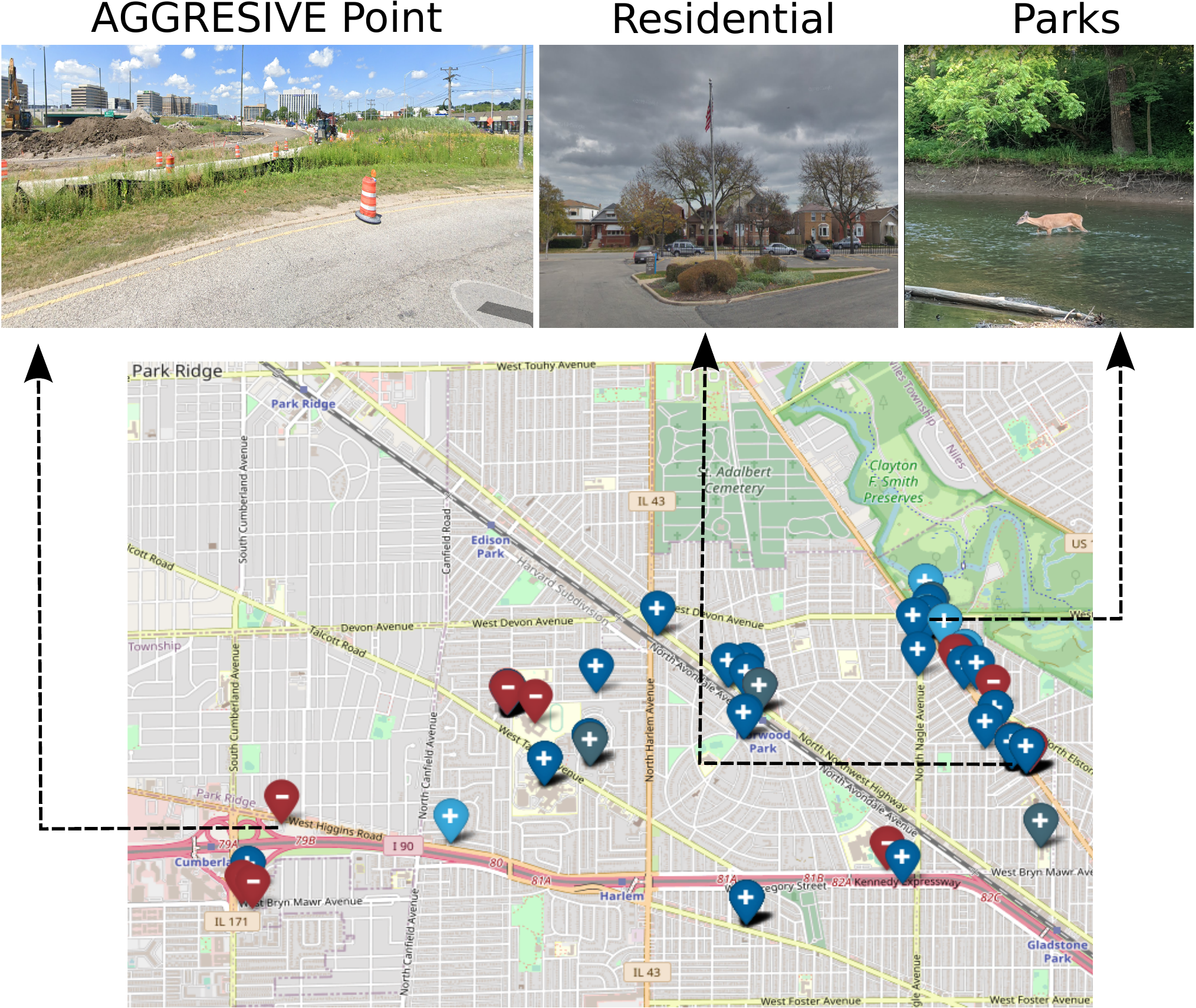}
   \caption{Perceptions map in Norwood Park. [Best in color]}
   \label{fig:chicago-norwood-perception-maps}
\end{figure}

In this section, we present the main characteristics of urban areas evaluated and map the spatial distribution of perceptions using markers, with ``Plus/Minus'' icons to represent the positive/negative perceptions. We also assigned colors to distinguish them: dark blue (GREAT); gray (LIVELY); blue (RESPECTFUL); light blue (SPECTACULAR); dark red (AGGRESSIVE); red (WRONG); light red (DEAD); and beige (CREEPY). 

The areas studied in Chicago are: Loop, Norwood Park and Wicker Park. Loop is the downtown area of Chicago, known as an important commercial and financial center of the city, attracting crowds of visitors and residents with different profiles to their several types of venues.

As we can see in Figure \ref{fig:chicago-loop-perception-maps}, Chicago's Downtown area contains a considerable concentration of distinct perceptions, suggesting that most of them coexist in the neighborhood. Such finding is not surprising because Downtown areas tend to present a wide variety of sounds, visual elements, odors, among other characteristics, which can potentially cause distinct perceptions to people. For this reason, different urban perceptions may be favored to occur.

Norwood Park is mostly residential, with some leisure area options, such as parks with sports fields and paths, and a hospital near to the central area of the neighborhood. As shown in Figure \ref{fig:chicago-norwood-perception-maps}, there are \emph{AGGRESSIVE} points spread in different parts of the neighborhood, which indicate that some people had bad experiences in this region. Beyond this perception, only \emph{GREAT}, \emph{LIVELY}, and \emph{RESPECTFUL} perceptions have occurred in the neighborhood during the considered period. Note also that we have comparatively fewer data concerning the other areas. Nevertheless, the proposed approach is still able to gather overall perceptions about this area.

\begin{figure}[!htp]
   \centering
   \includegraphics[scale=0.45]{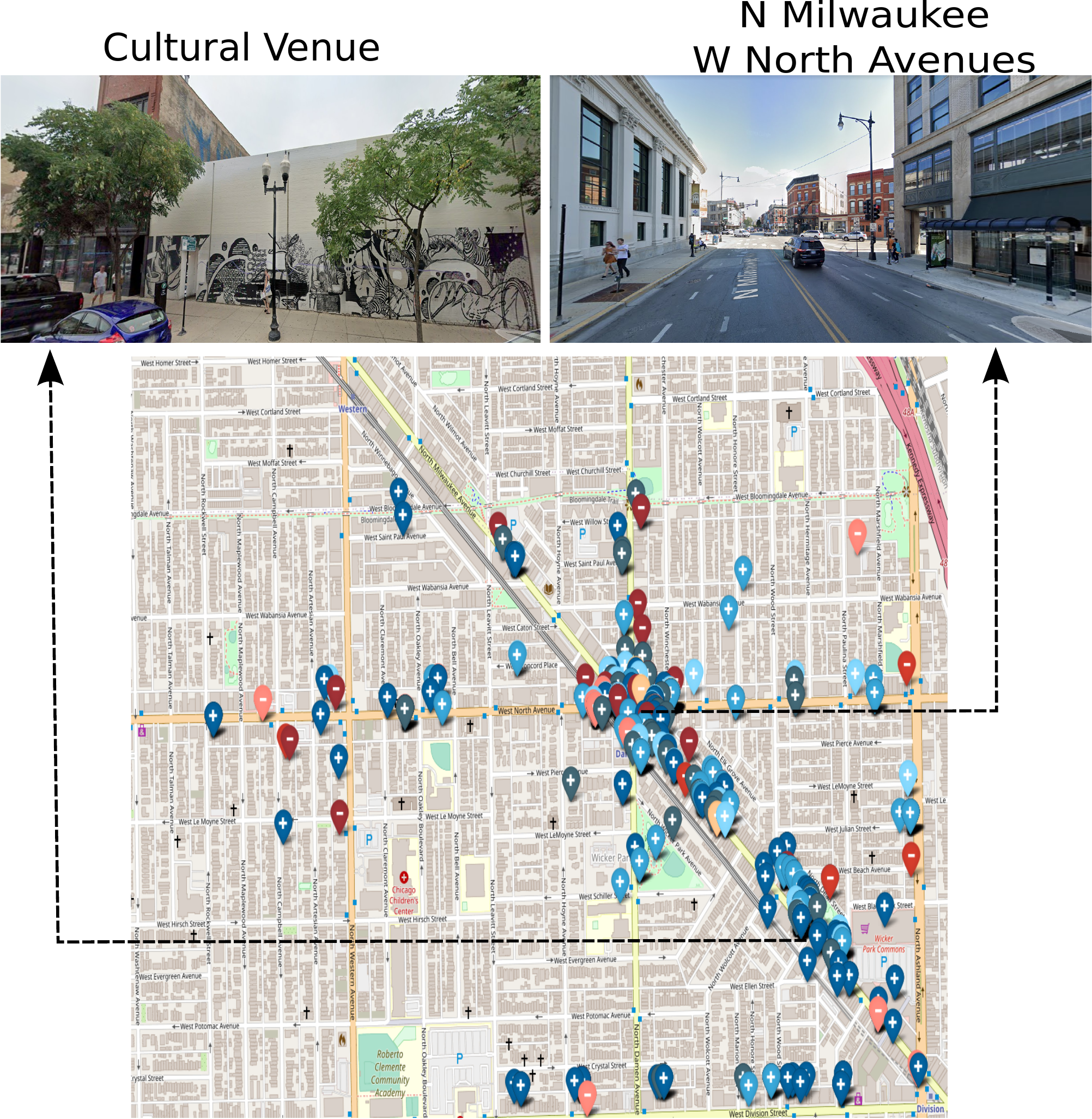}
   \caption{Perceptions map in Wicker Park. [Best in color]}
   \label{fig:chicago-wicker-perception-maps}
\end{figure}

%NYC
\begin{figure}[!htp]
   \centering
   \includegraphics[scale=0.35]{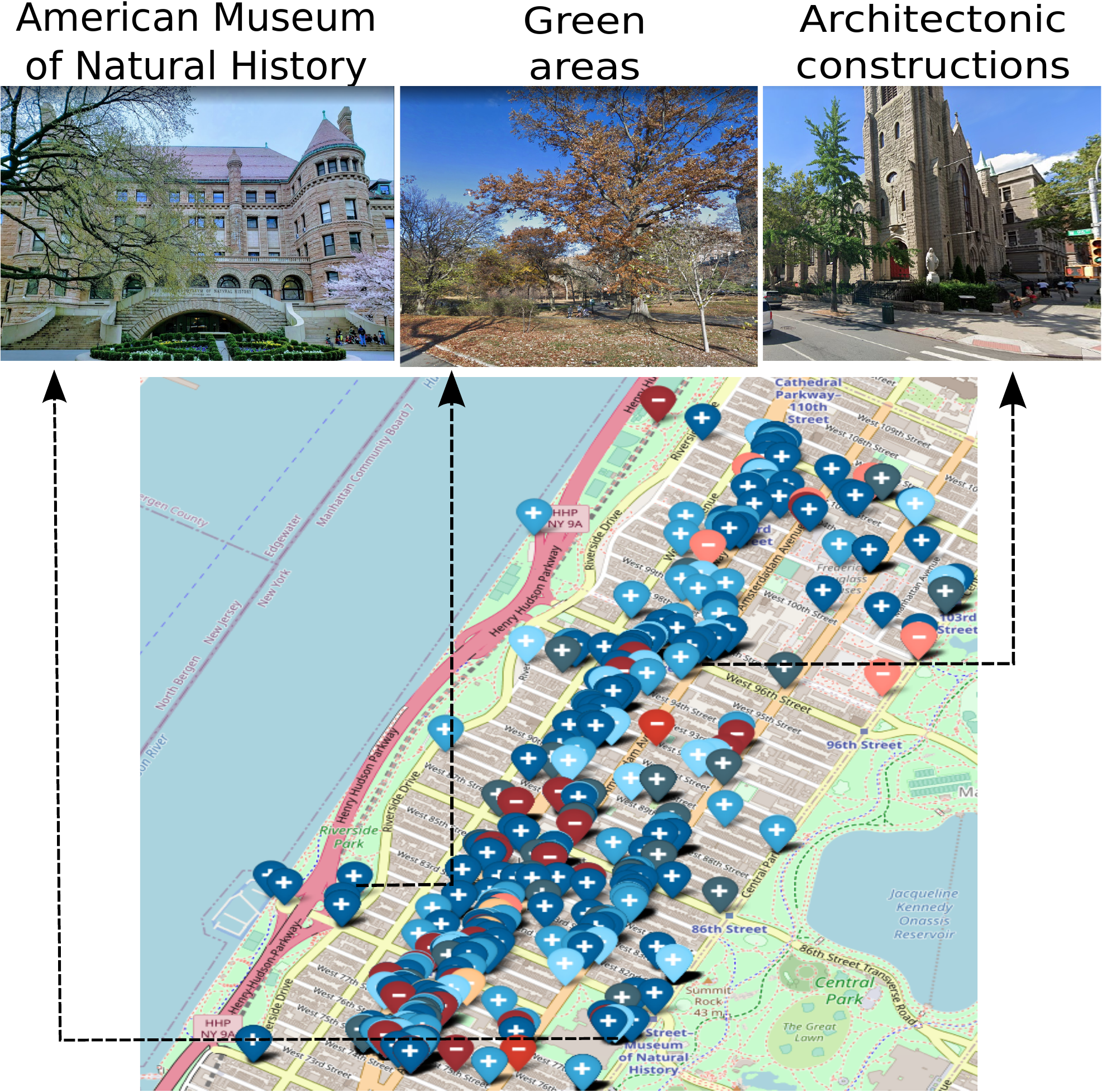}
   \caption{Perceptions map in Upper West Side. [Best in color]}
   \label{fig:ny-upperwest-perception-maps}
\end{figure}

Considering the Wicker Park region, known as a hub for shopping, eating, and cultural activities in the city, we can see in Figure \ref{fig:chicago-wicker-perception-maps} several overlaps among the perceptions in the intersection between the ``N Milwaukee'' and ``W North'' avenues, where restaurants and bars are concentrated. This indicates that, for example, shopping or eating in these areas is an everyday activity to be performed, attracting a large number of citizens and visitors, inducing a mix of perceptions. We can also see that another interesting finding in this neighborhood is located close to Dean Playground Park. In this spot, there are cultural venues, favoring positive perceptions.

Studying now NYC, the three regions in our evaluation are: Upper West Side, Park Slope and Gowanus, and Jamaica. Upper West Side is famous for the architectonic constructions. Besides, it offers several cultural venues, such as the American Museum of Natural History. It also concentrates an academic community due to its proximity to Columbia University, and has several options for gastronomy and nightlife. The perceptions in the Upper West Side, Figure \ref{fig:ny-upperwest-perception-maps}, are concentrated around the Broadway and Amsterdam Avenues, and also spread by the neighborhood close to the shopping and green areas. The mainly perceptions are \emph{GREAT}, \emph{SPECTACULAR} and \emph{RESPECTFUL}.

\begin{figure}[!htp]
   \centering
   \includegraphics[scale=0.35]{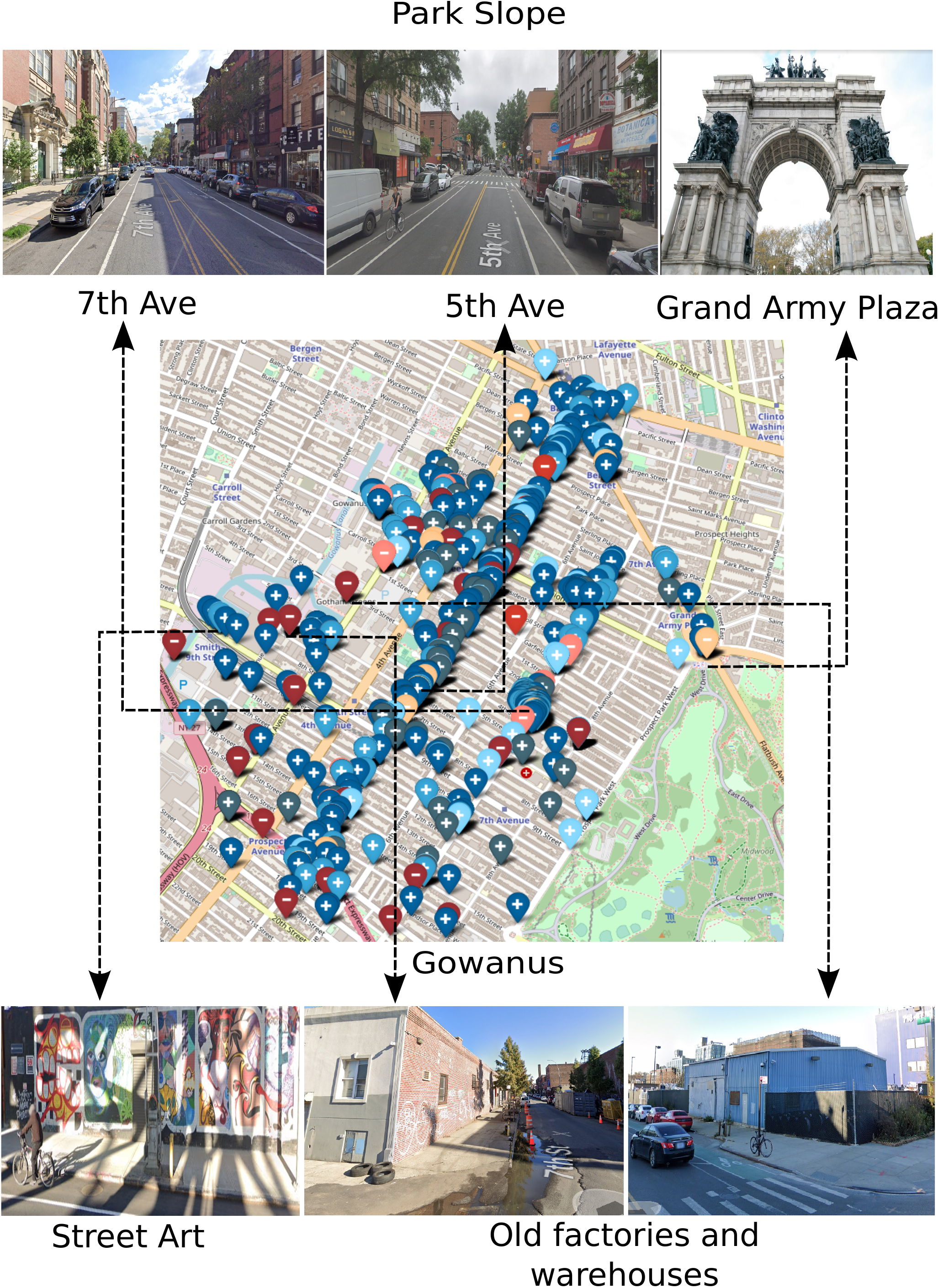}
   \caption{Perceptions map in Park Slope and Gowanus. [Best in color]}
   \label{fig:ny-parkslope-perception-maps}
\end{figure}

\begin{figure}[!htp]
   \centering
   \includegraphics[scale=0.4]{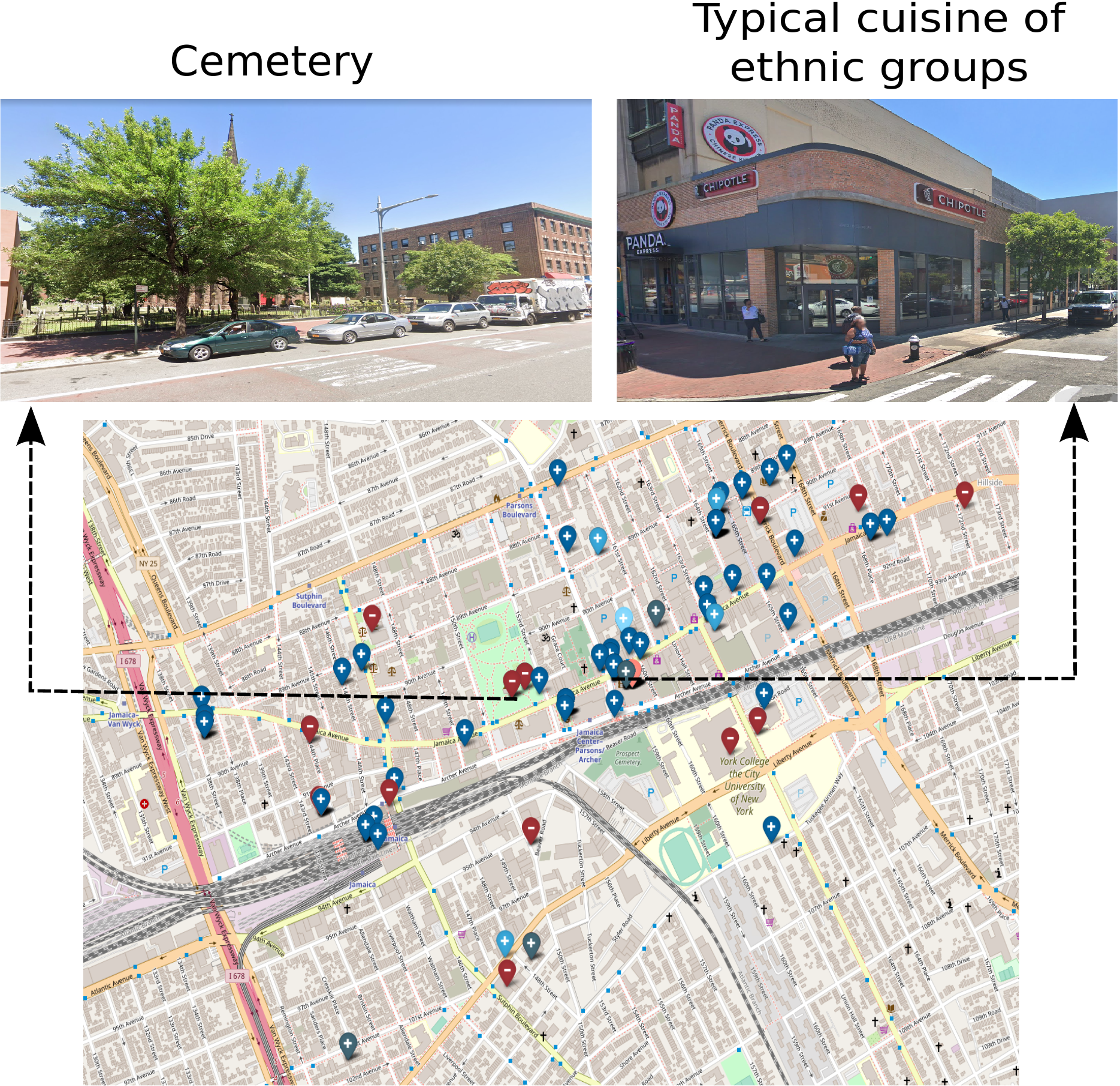}
   \caption{Perceptions map in Jamaica. [Best in color]}
   \label{fig:ny-jamaica-perception-maps}
\end{figure}

The Park Slope and Gowanus are located in Brooklyn and, despite the close geographic proximity, they have very distinct characteristics, as shown in Figure \ref{fig:ny-parkslope-perception-maps}. Park Slope is famous for the charming buildings, great food establishments and lovely parks, such as Prospect Park and Grand Arm Plaza, which favor to emerge several positive perceptions, mainly the surrounding of \nth{5} Avenue, Union Street, Barclays Center and along \nth{7} Avenue. On the other hand, Gowanus is an industrial business zone with a few residential areas mixed among old factories and warehouses, where there are several points of perception \emph{AGGRESSIVE}. Recently, Gowanus has been attracting artists to the neighborhood, who have made a significant change to the area, such as street art murals and vibrant bars.

%London
\begin{figure}[!htp]
   \centering
   \includegraphics[scale=0.35]{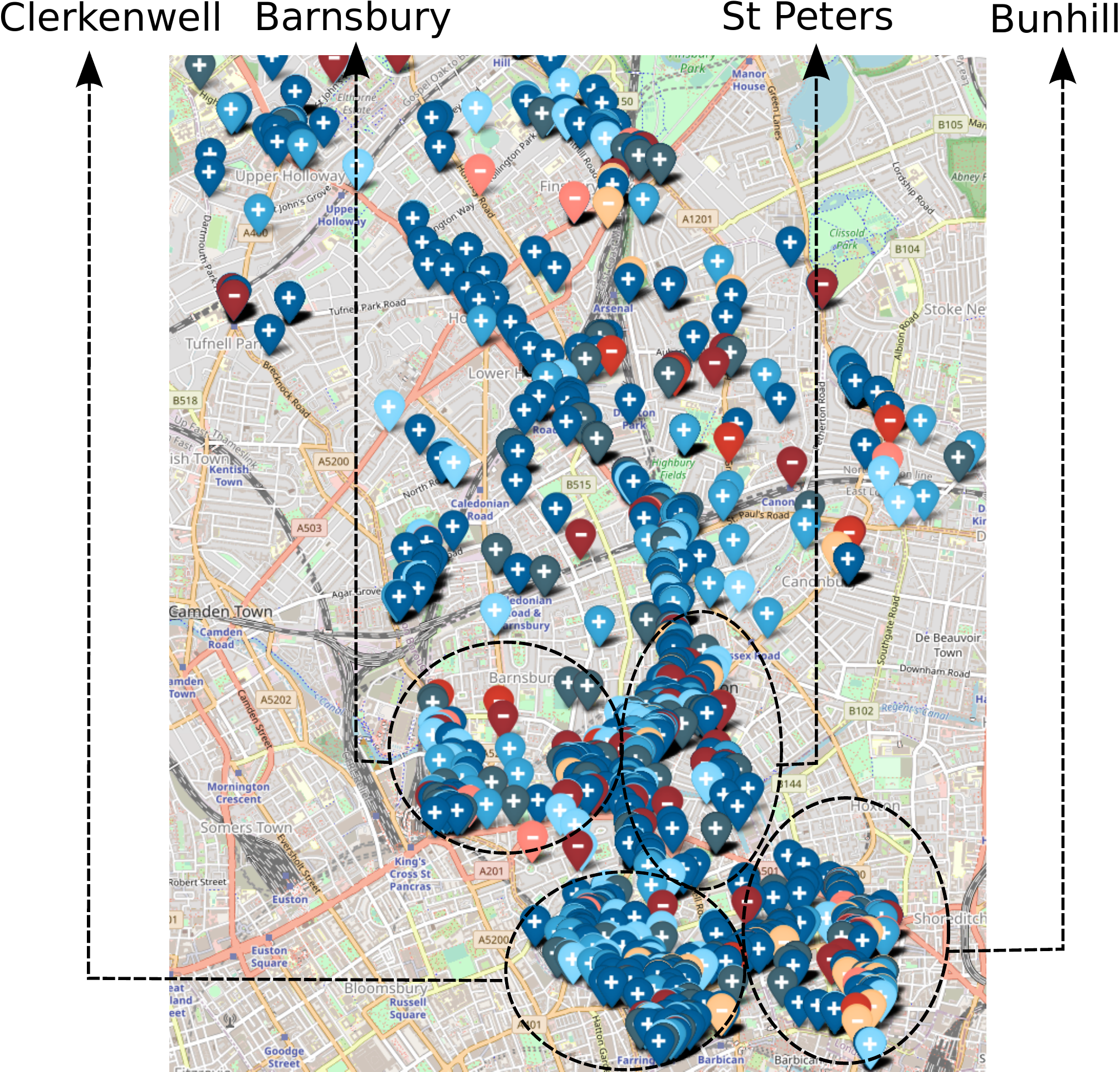}
   \caption{Perceptions map in Islington. [Best in color]}
   \label{fig:london-islington-perception-maps}
\end{figure}

Situated in the Queens borough, Jamaica is a residential neighborhood that is mainly populated with Latinos \citep{jamaica2018}. Other minority populations, such as Asians and African Americans, also live in large numbers in this area \citep{jamaica2018}. Due to the high diversity of the community living in the neighborhood, there are many restaurants and grocery stores with typical cuisine of ethnic groups, arising positive perceptions \emph{GREAT}, \emph{LIVELY} and \emph{RESPECTFUL} close to those places, as shown in Figure \ref{fig:ny-jamaica-perception-maps}. By contrast, Jamaica suffers from a high rate of violent crimes \citep{jamaica2018}, being a negative aspect of the neighborhood, where several points of \emph{AGGRESSIVE} are spread in the area.

Finally, we evaluated London Boroughs rather than neighborhoods, since most neighborhoods have small areas, where most of them have insufficient data to perform the analysis. The selected regions for evaluating are Islington, Hammersmith and Fulham, and Lambeth. Islington concentrates diverse perception across its area, as shown in Figure \ref{fig:london-islington-perception-maps}. In that figure, we can also see that the positive ones stand out. Among Islington districts, we can highlight Bunhill, Clerkenwell, Barnsbury, and Saint Peters, as the districts that concentrate the highest number of perceptions. This fact might be motivated by the proximity of these districts with London's downtown, i.e., which implies a higher diversity and a higher number of people.

\begin{figure}[!htp]
   \centering
   \includegraphics[scale=0.35]{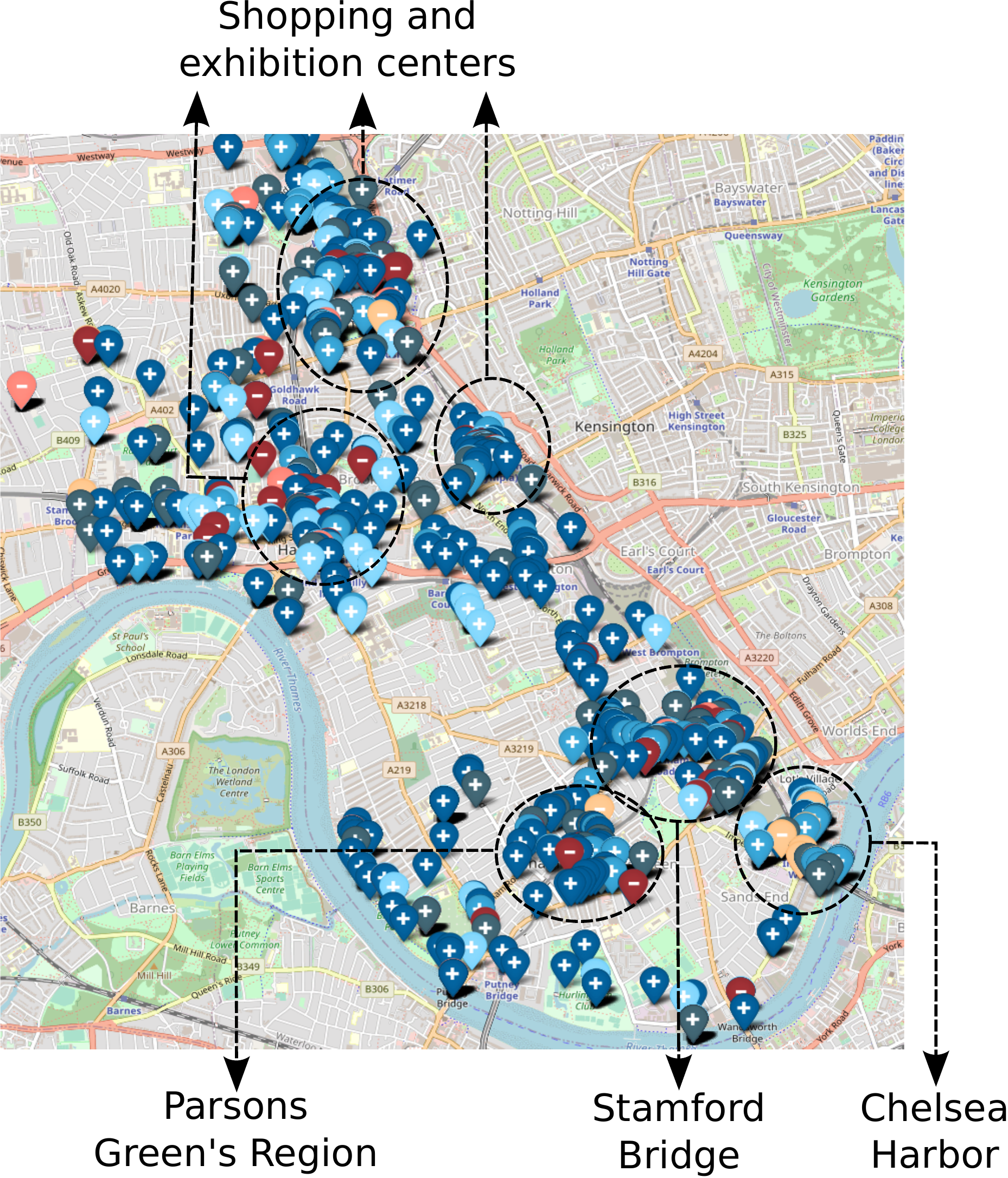}
   \caption{Perceptions map in Hammersmith and Fulham. [Best in color]}
   \label{fig:london-hammersmith-perception-maps}
\end{figure}

\begin{figure}[!htp]
   \centering
   \includegraphics[scale=0.45]{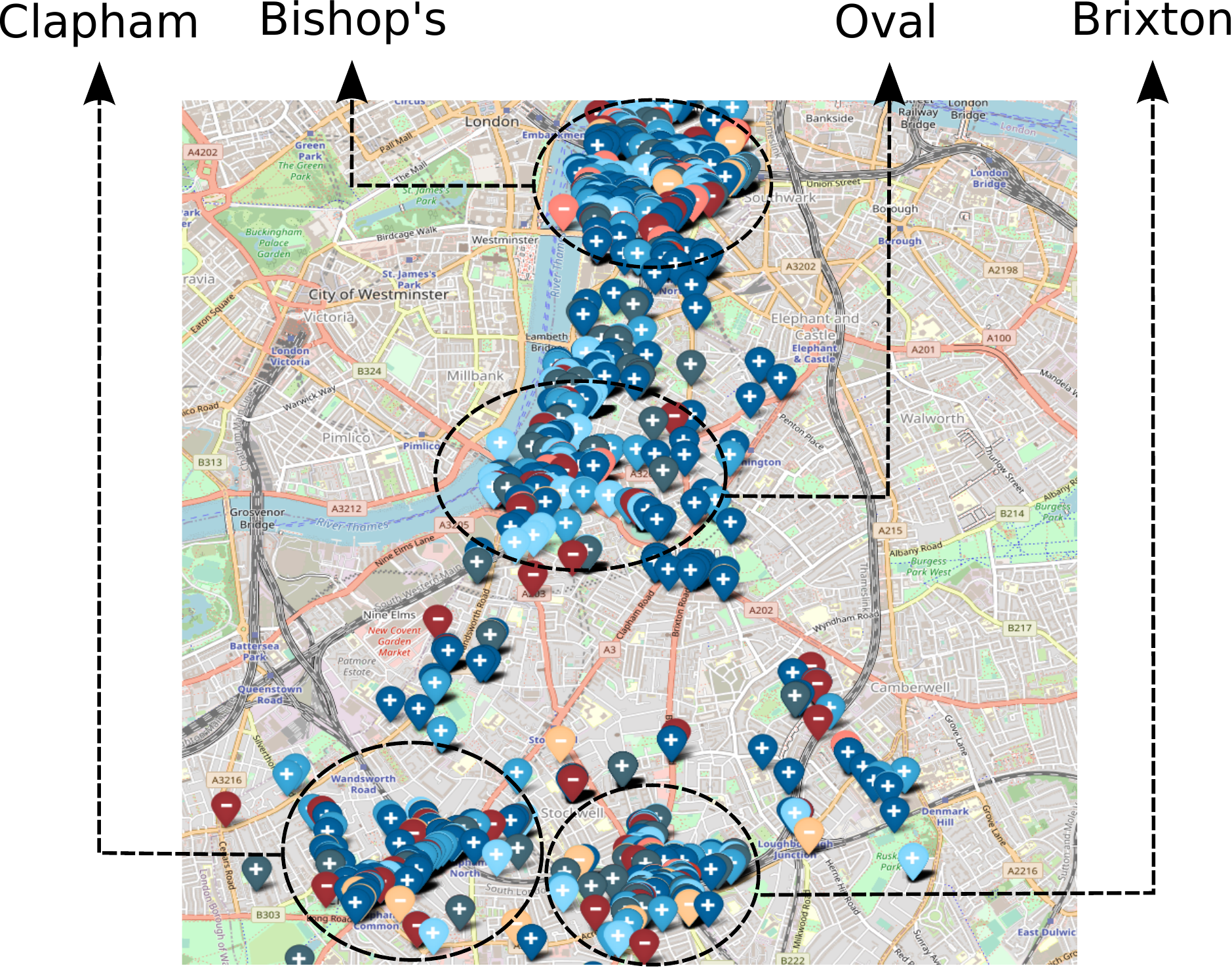}
   \caption{Perceptions map in Lambeth. [Best in color]}
   \label{fig:london-lambeth-perception-maps}
\end{figure}

With one of the most expensive places to live in London, Hammersmith and Fulham have been attracting affluent families and prosperous young people to live there. Moreover, Hammersmith and Fulham house old-fashion pubs, soccer stadiums, cultural venues, shops, cafes, and malls, being a popular place to visit in the city. In Hammersmith and Fulham, we can see in Figure \ref{fig:london-hammersmith-perception-maps}, close to Stamford Bridge, several points of positive perceptions, mainly \emph{GREAT}, \emph{LIVELY}, and \emph{RESPECTFUL} categories, mixed with points of the perception \emph{AGGRESSIVE}, the negative one. Moreover, there are dense groups of positive category points in various parts of the borough. For instance, in Parsons Green and Chelsea Harbour in Fulham, and Shopping centers and exhibition centers in Hammersmith, indicating that these areas are desirable to visit. We can also see some points of categories \emph{AGGRESSIVE} and \emph{CREEPY} close by and, in some cases, overlapping those groups, suggesting, probably, problems inherent to crowded places.

Turning our attention to Lambeth, as we can see in Figure \ref{fig:london-lambeth-perception-maps}, there is a concentration of perceptions in the following districts: Bishop's, Oval, Brixton, and Clapham. Besides, there are several groups of mixed perceptions spread in the remaining areas of Lambeth. Despite common unbalance in the number of perceptions in other scenarios, mainly with \emph{GREAT} perception, almost all perceptions appear in a significant quantity in this region.

\subsection{Temporal Analysis at Neighborhood Scale}
\label{subsec:temporal_analysis}
To evaluate the urban perceptions extracted from the studied areas, we conducted a temporal analysis to measure the persistence of perceptions identified over time. Such analysis is useful for ranking the perceptions in the neighborhoods, as well as to identify if perceptions keep unchanged independent of the period. To this end, we calculate the z-scores similar to some authors, such as \citep{aiello2016chatty, quercia2016emotional, redi2018spirit}, defined as follow:  

\begin{equation}\label{eq:perc_strength}
    \text{z-score}_{i}^j (n)= \frac{X_{i}^j(n) - \mu(X_{i}^j)}{\sigma(X_{i}^j)},
\end{equation}
where $i$ denotes one specific perception category among all we consider, $j$ denotes the period (here we consider months), and $n$ denotes a specific city neighborhood among all neighborhoods. In addition, $X_{i}^j(n)$ indicates the number of points of perception $i$ in the neighborhood $n$ during the month $j$, whereas $\mu(X_{i}^j)$ and $\sigma(X_{i}^j)$ are the mean number and the standard deviation of points of perception $i$ during the month $j$ in every neighborhood of the city, respectively. Using z-score, we can capture, for every neighborhood, the perceptions that stand out, even when the number of samples of perception categories is very unbalanced, which is our case. For instance, the category \emph{GREAT} is by far the most numerous, and the reason for that could be diverse, e.g., people tend to share more the ``good'' moments in social media. 

\begin{figure}[!htp]
   \centering
    \subfloat[Loop (Downtown).]{
        \includegraphics[width=0.35\textwidth]{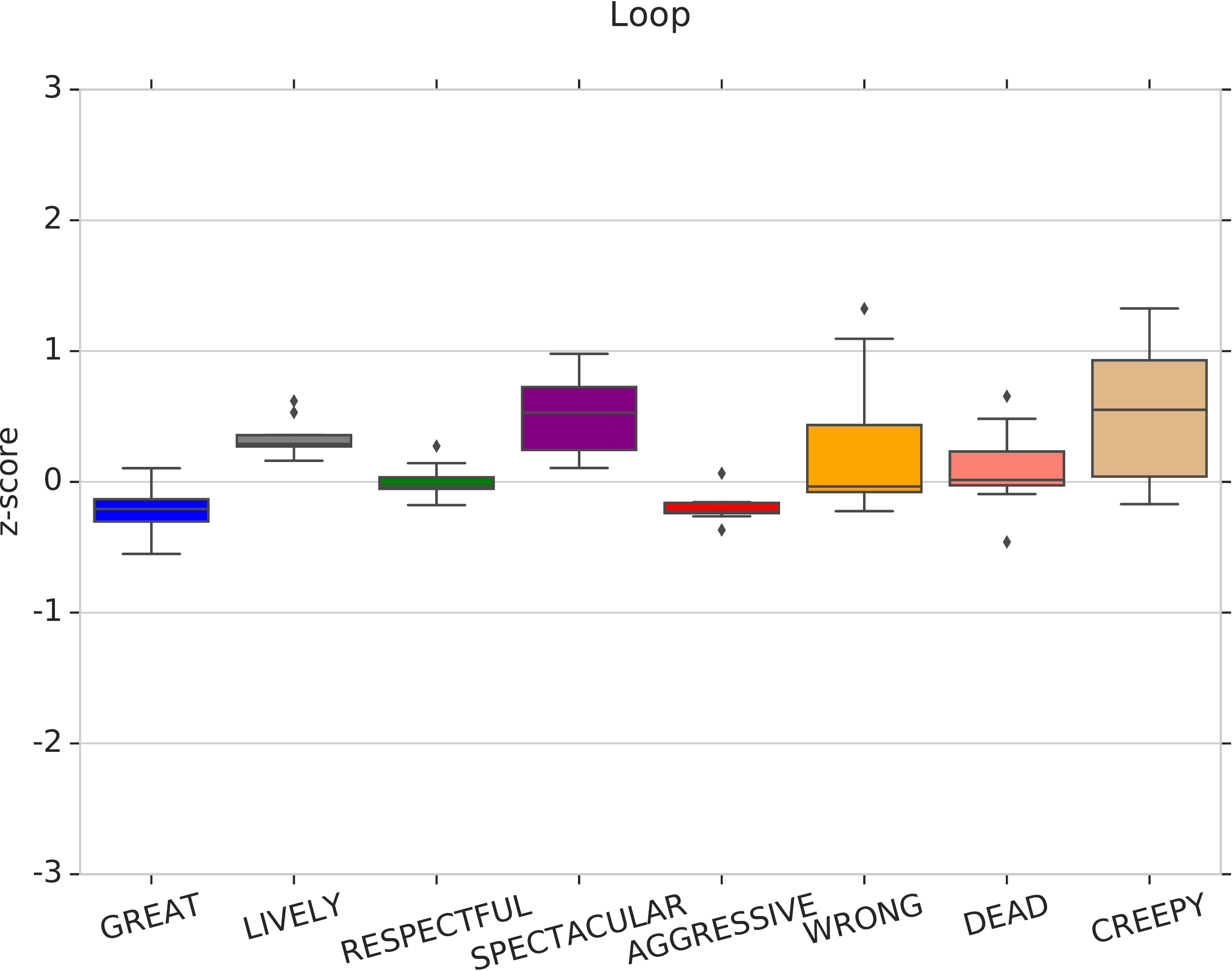}} \hspace{0.1cm}
    \subfloat[Wicker Park.]{
        \includegraphics[width=0.35\textwidth]{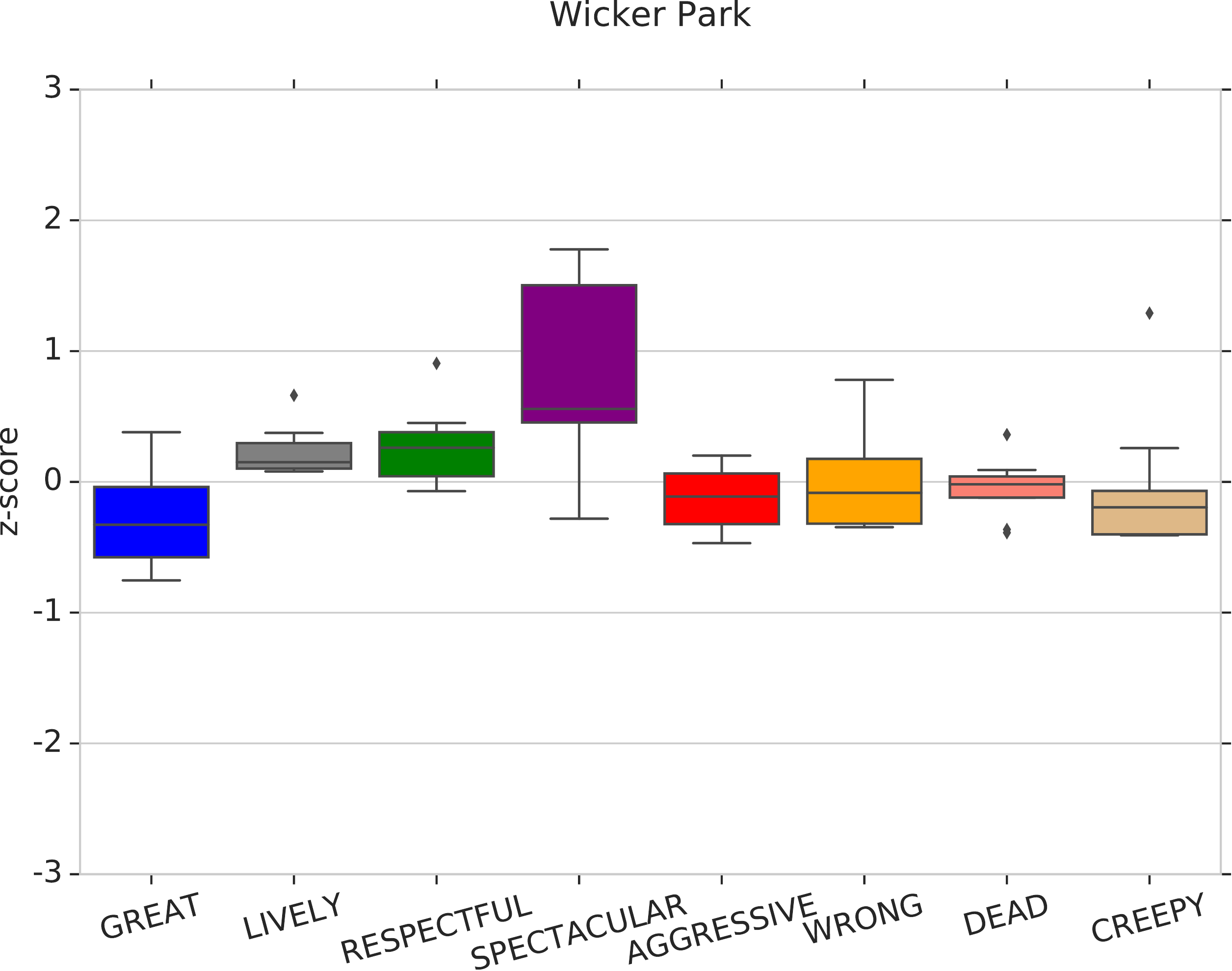}} \hspace{0.1cm}
    \subfloat[Norwood Park.]{
        \includegraphics[width=0.35\textwidth]{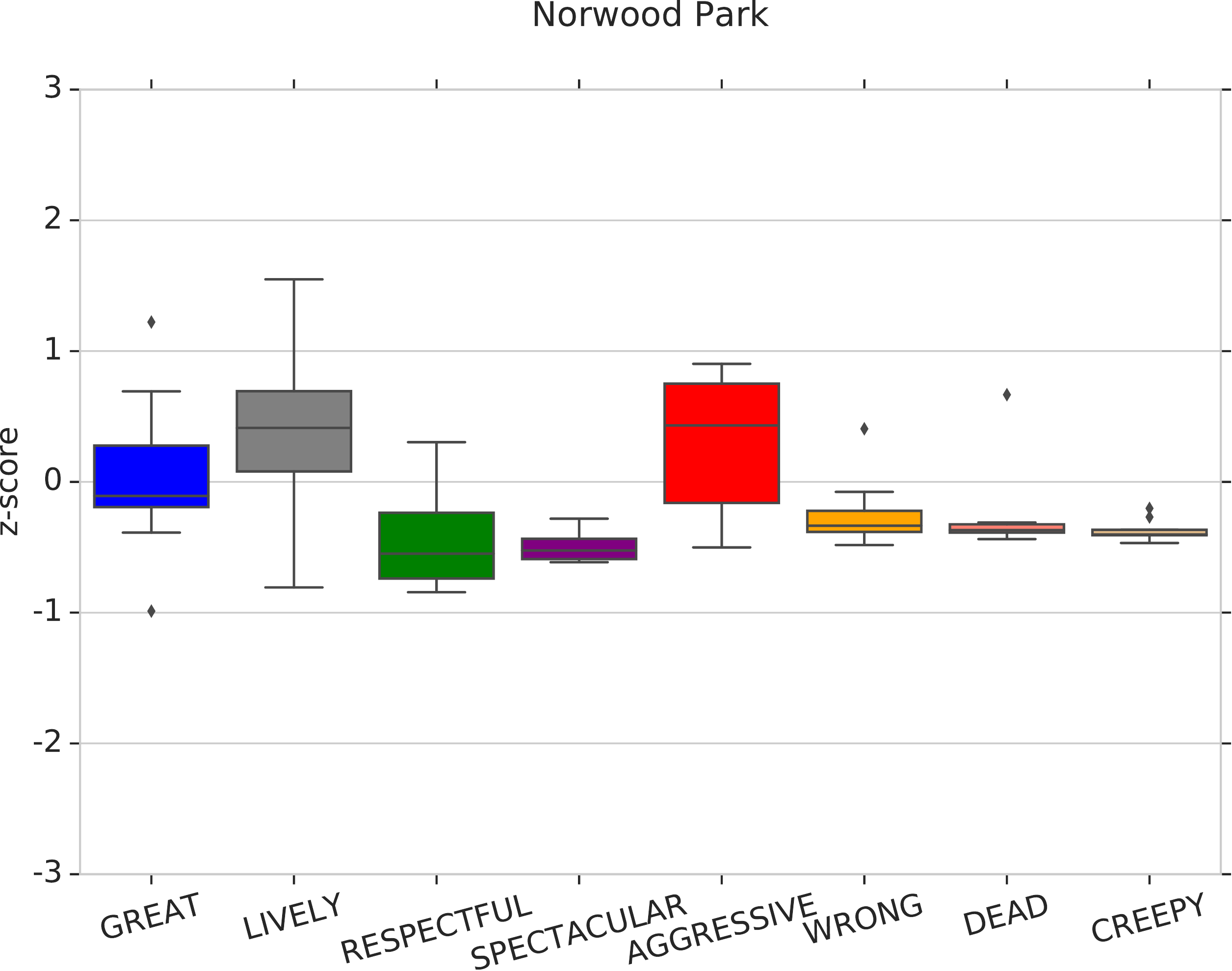}}
    \caption{Boxplots containing the perception strength of Chicago's neighborhoods.}
    \label{fig:chicago-z-scores}
\end{figure}

%Chicago
Figure \ref{fig:chicago-z-scores} shows the results (exploring boxplots) for Chicago. As we can see in Figure \ref{fig:chicago-z-scores}(a), several perceptions occurred in the Downtown area with low variation along the time (less than one standard deviation in most cases) and similar intensity (median $z$-score near to $0$). Among them, \emph{LIVELY}, \emph{SPECTACULAR}, and \emph{CREEPY} are the most significant ones and tend to occur with high strength in this area. \emph{WRONG} can appear with high intensity on few months ($z$-score $\geq 1$); however, note that the median is close to $0$. This is not the case for \emph{CREEPY} perception, indicating that visitors recurrently had some issues during their personal experience when visiting the neighborhood, generating this class of perception. 

Wicker Park is predominantly \emph{SPECTACULAR} with the higher median perception strength, as shown in Figure \ref{fig:chicago-z-scores}(b). Moreover, \emph{LIVELY} and \emph{RESPECTFUL} also have a significant occurrence in the neighborhood. At the same time, the remaining perceptions keep the median perception strength below the expected to a neighborhood ($z$-score $<0$). The perception about the region of Norwood Park neighborhood is predominantly \emph{LIVELY}, followed by \emph{AGGRESSIVE}. Note that for other perceptions, the median strength is below $0$, indicating they are hardly related to this area.

\begin{figure}[!htp]
   \centering
    \subfloat[Upper West Side.]{
        \includegraphics[width=0.35\textwidth]{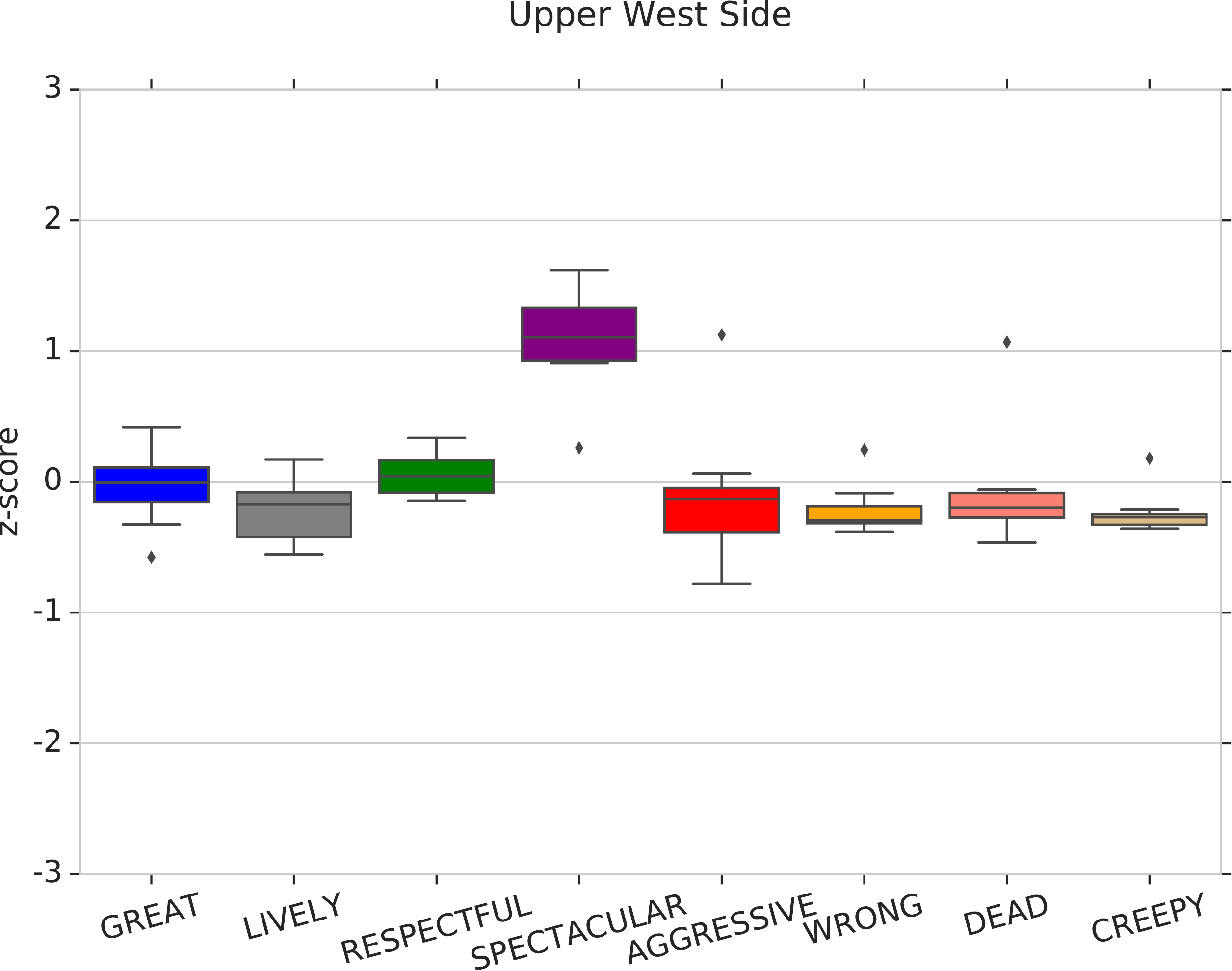}} \hspace{0.2cm}
    \subfloat[Park Slope and Gowanus.]{
        \includegraphics[width=0.35\textwidth]{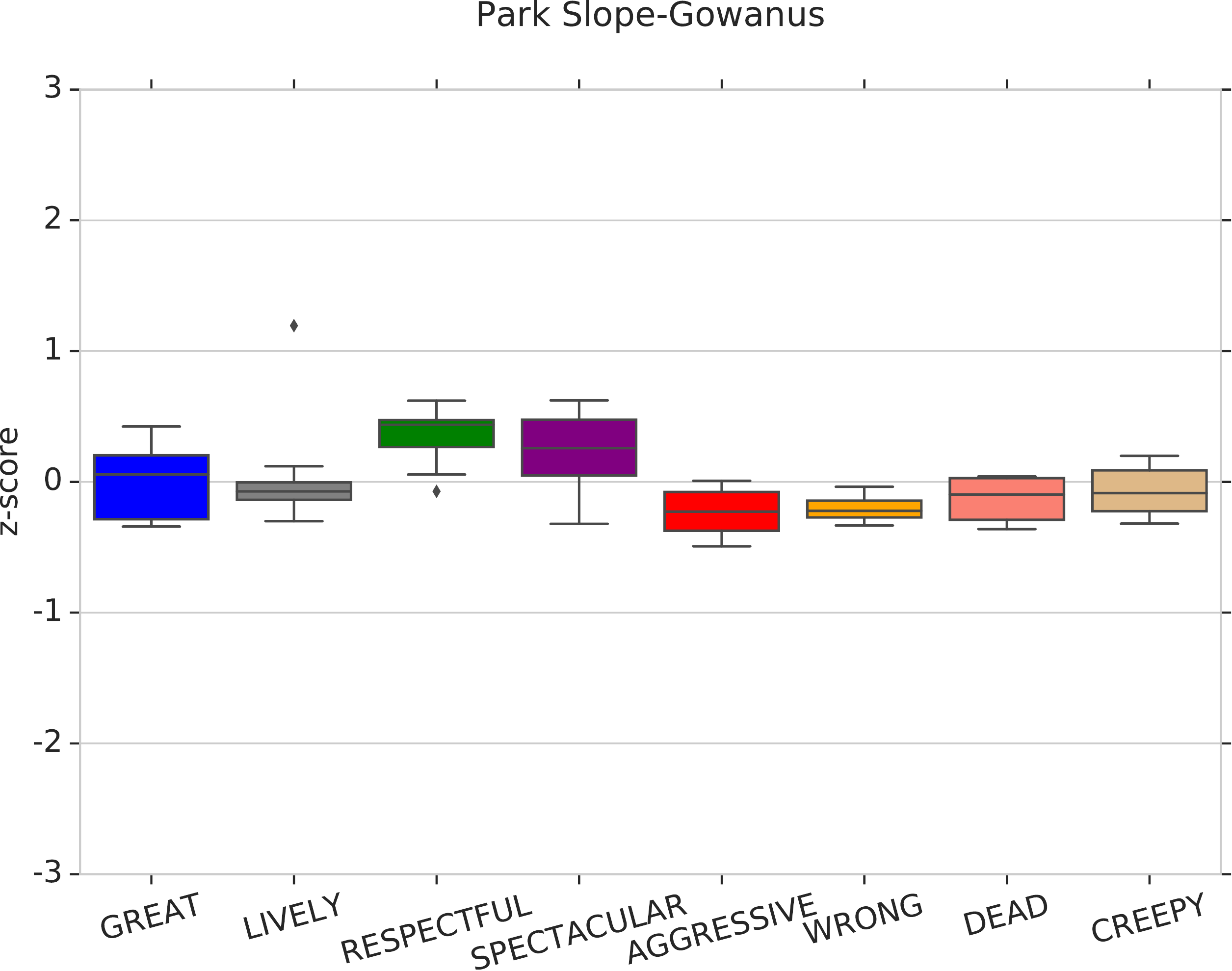}} \hspace{0.2cm}
    \subfloat[Jamaica.]{
        \includegraphics[width=0.35\textwidth]{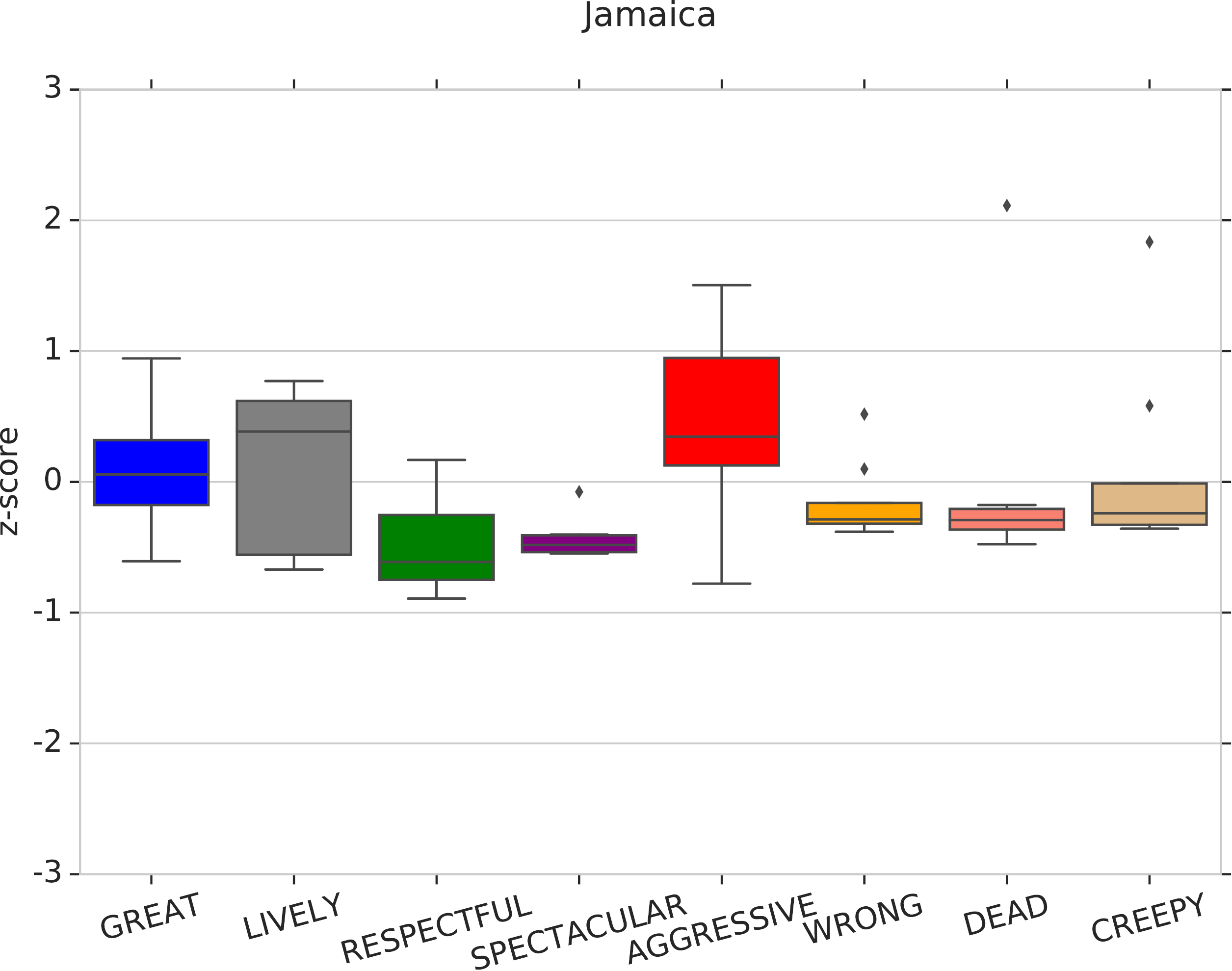}}
    \caption{Boxplots containing the perception strength of NYC's neighborhoods.}
    \label{fig:ny-z-scores}
\end{figure}

%NYC
Turning our attention to NYC, Figure \ref{fig:ny-z-scores} shows the results for its studied neighborhoods. As shown in Figure \ref{fig:ny-z-scores}(a), the Upper West Side region is mainly \emph{SPECTACULAR} (median $z$-score $\geq 1$), but also is \emph{GREAT} and \emph{RESPECTFUL}. Negative perceptions have low strength overall. Studying Park Slope and Gowanus neighborhoods, Figure \ref{fig:ny-z-scores}(b), most perceptions kept stable throughout the months. \emph{GREAT}, \emph{RESPECTFUL} and \emph{SPECTACULAR} are the most intense ones in the area. For this area, negative perceptions are also not significant (median $z$-score $< 0$). As we can see in Figure \ref{fig:ny-z-scores}(c), the median perception strength of \emph{LIVELY} and \emph{AGGRESSIVE} is the most predominant in Jamaica neighborhood. As we can see, \emph{GREAT} tends to occur moderately in this region.

\begin{figure}[!htp]
   \centering
    \subfloat[Hammersmith and Fulham.]{
        \includegraphics[width=0.35\textwidth]{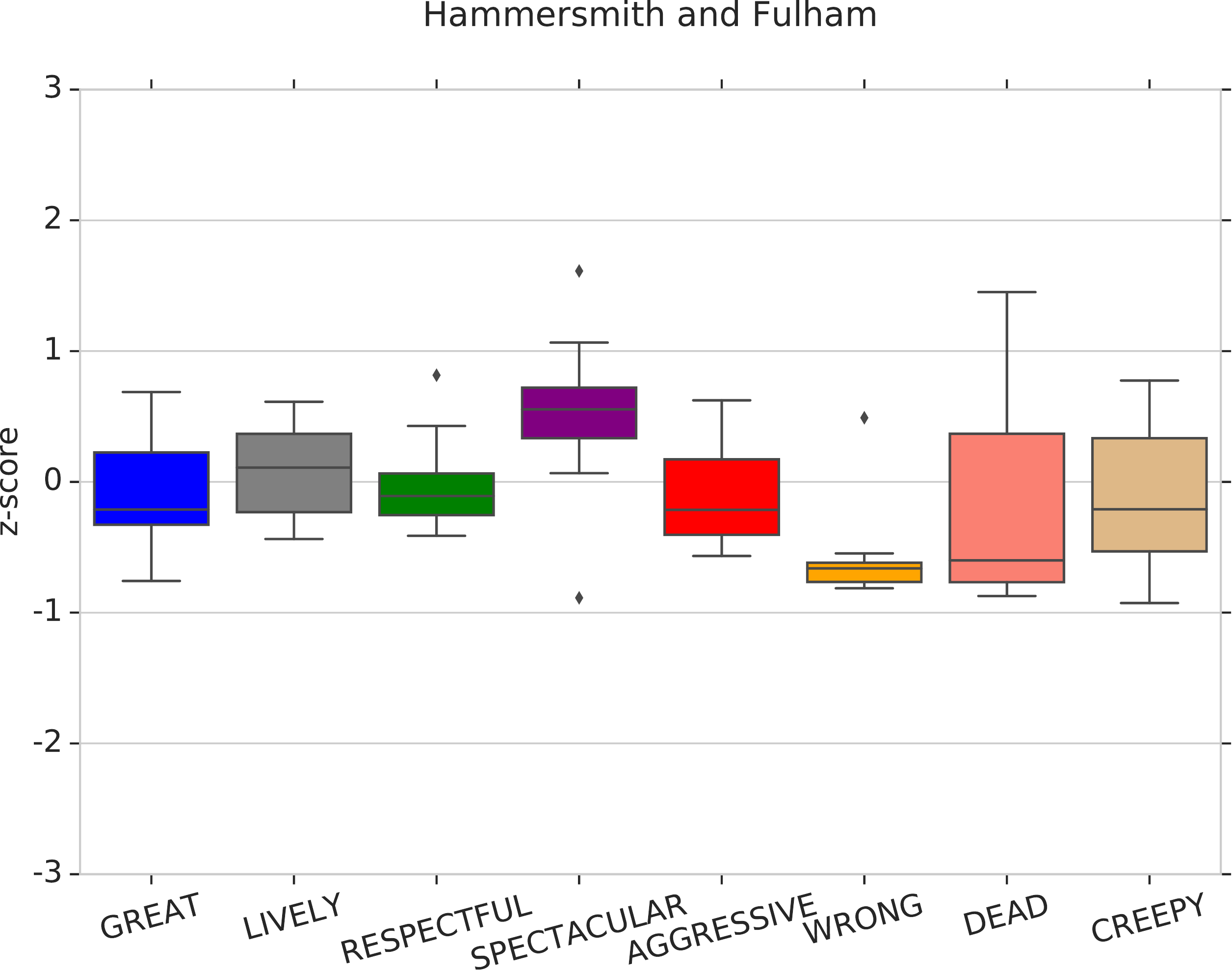}} \hspace{0.2cm}
    \subfloat[Islington.]{
        \includegraphics[width=0.35\textwidth]{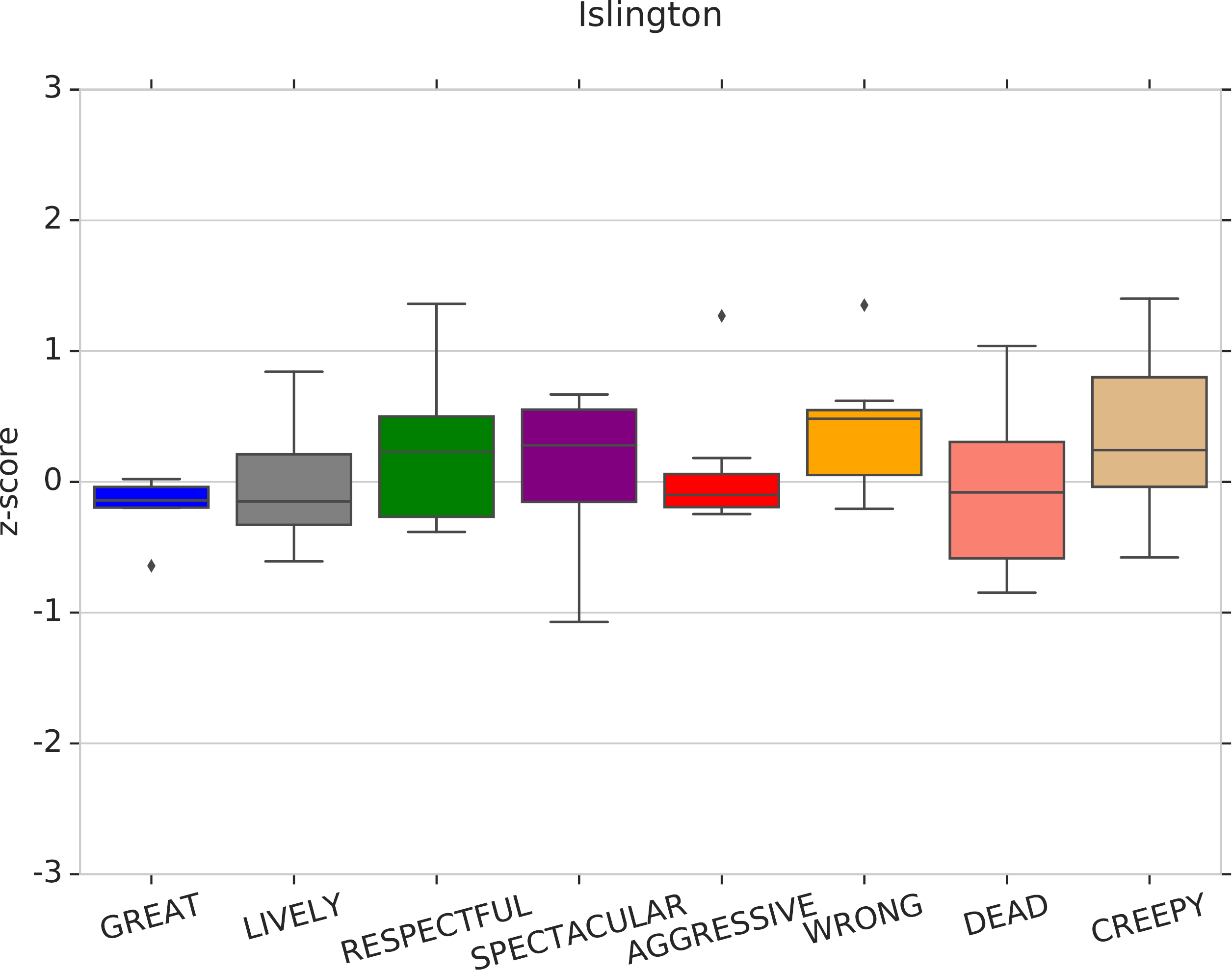}} \hspace{0.2cm}
    \subfloat[Lambeth.]{
        \includegraphics[width=0.35\textwidth]{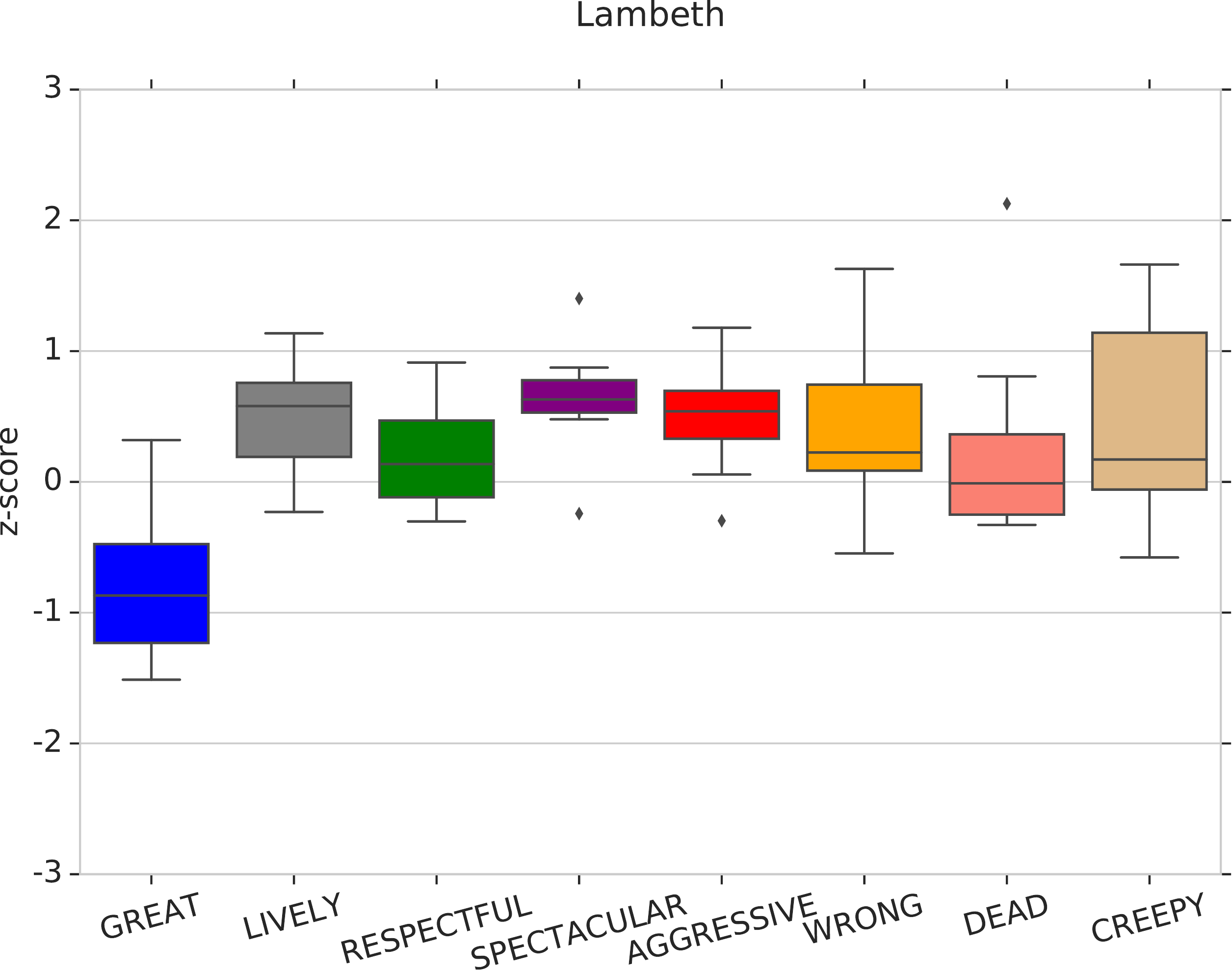}}
    \caption{Boxplots containing the perception strength of London's neighborhoods.}
    \label{fig:london-z-scores}
\end{figure}

%London
Finally, Figure \ref{fig:london-z-scores} shows the urban perceptions for London. It is possible to see that Hammersmith and Fulham (Figure \ref{fig:london-z-scores}(a)) has \emph{LIVELY} and \emph{SPECTACULAR} as predominant perceptions. Note that the perception \emph{RESPECTFUL} could be considered a secondary perception to this region, with the $z$-score value close to zero and low variation. Other perceptions have low $z$-score values, mainly the negative ones, being perceptions less relevant. For Islington, Figure \ref{fig:london-z-scores}(b), we observe an equilibrium among positive and negative perceptions. \emph{RESPECTFUL} and \emph{SPECTACULAR} are the most significant positive perceptions, and \emph{WRONG} and \emph{CREEPY} are the negative perceptions more notable. Lambeth borough shows the most heterogeneous behavior among the studied areas (Figure \ref{fig:london-z-scores}(c)), where six of the eight perceptions captured in this area have high $z$-score values. We can see that \emph{LIVELY} and \emph{SPECTACULAR}, followed by \emph{AGGRESSIVE} and \emph{CREEPY}, are the most relevant in this area.

\subsection{Comparison with Place Pulse 2.0}
\label{subsec:compare_PPulse}
To verify if the perception identified by our method reflects the overall people's opinion, we have conducted a comparative analysis with results obtained by the research project Place Pulse 2.0 \citep{dubey2016deep}. This project collected volunteers' perception based on features present in urban outdoor images, containing more than 1,2{M} comparisons between pairs of images. According to volunteers' opinions, the comparison captures which image better matches with a particular perception regarding safety, wealthy, beautiful, lively, boring, and depressing. More information about the process can be obtained in \citep{dubey2016deep}. The researchers collected opinions based on images of 56 cities, among them Chicago, NYC, and London (responses used in this study). 

\begin{figure}[!htp]
  \centering
    \subfloat[Chicago]{
        \includegraphics[width=0.3\textwidth]{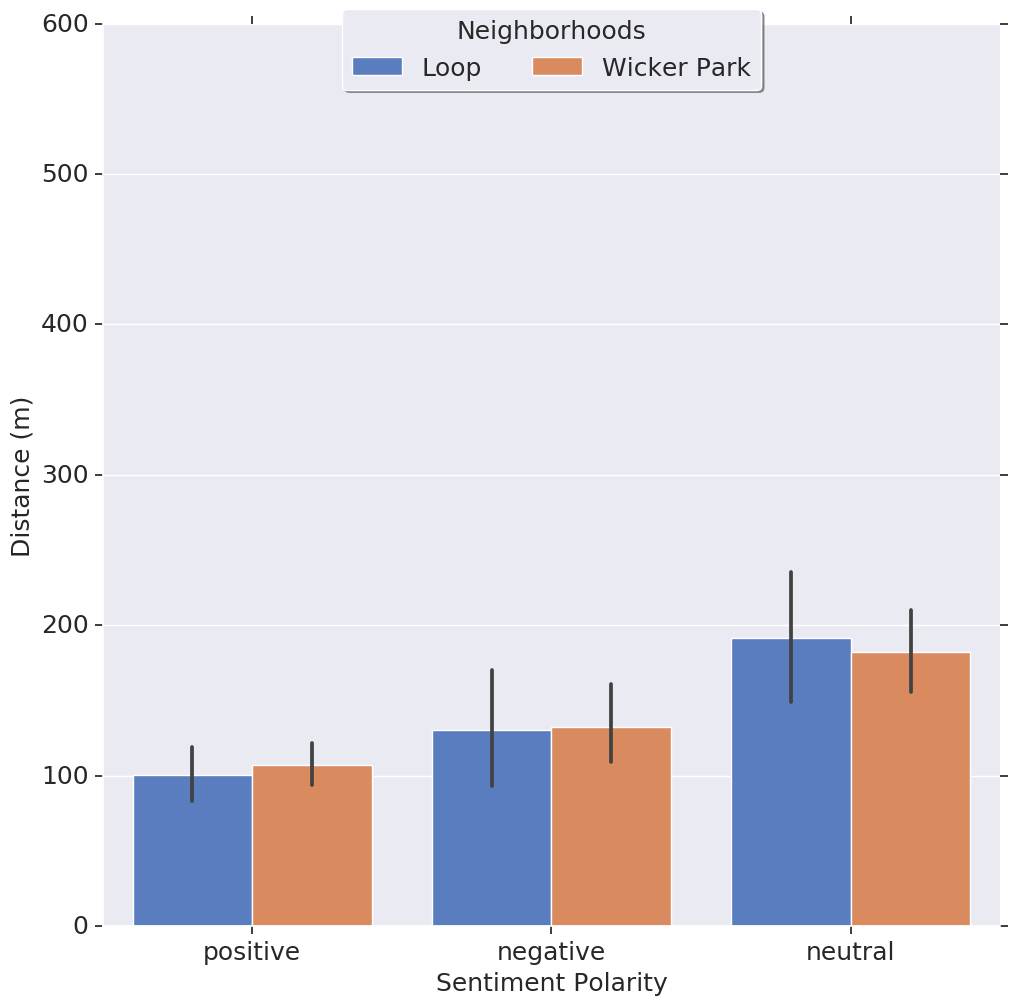}} \hspace{0.2cm}
    \subfloat[NYC.]{
        \includegraphics[width=0.3\textwidth]{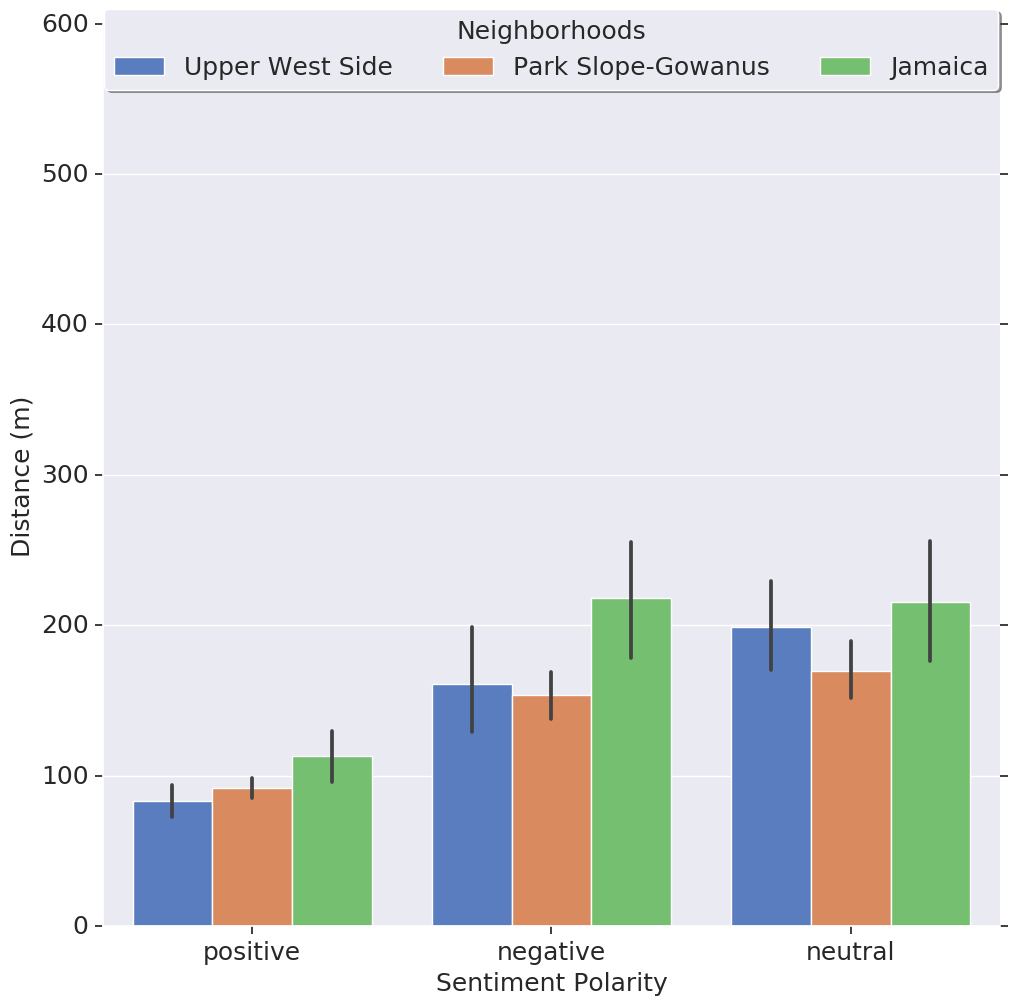}}\hspace{0.2cm}
    \subfloat[London.]{
        \includegraphics[width=0.3\textwidth]{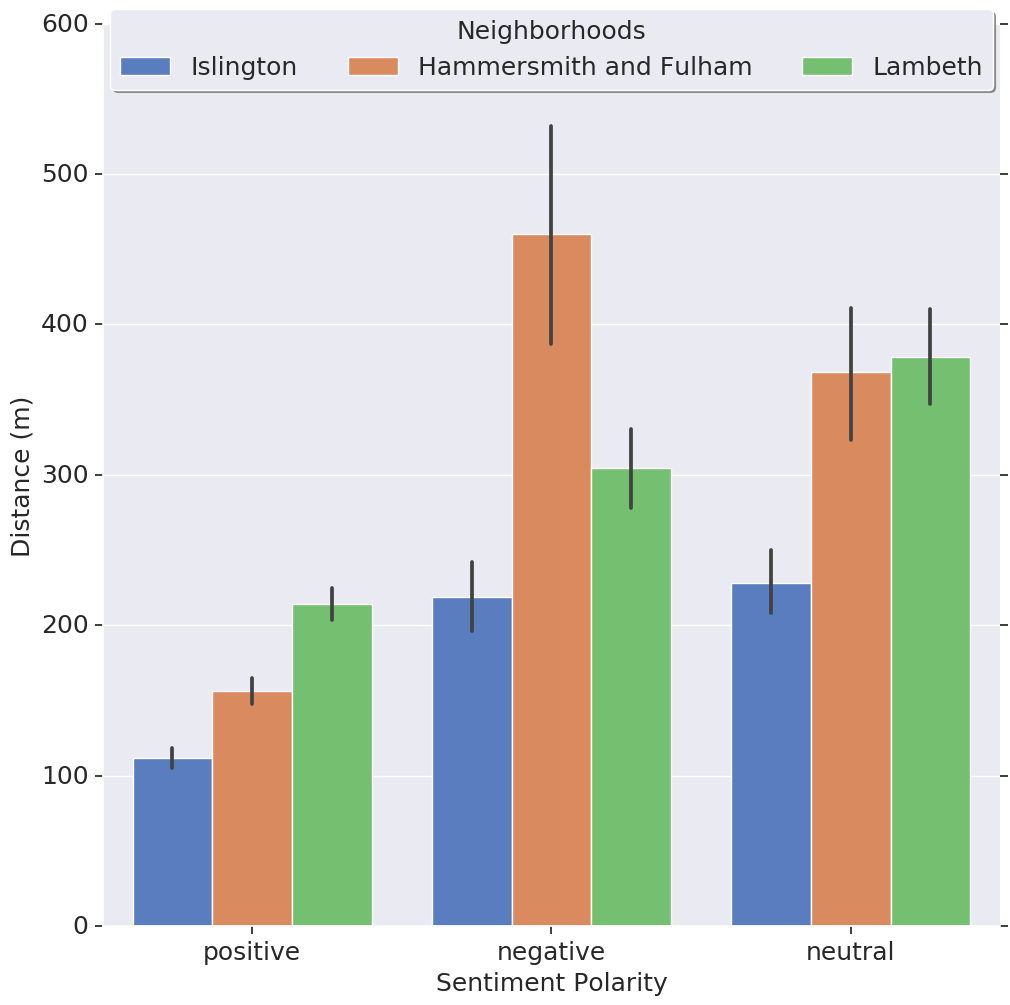}}
    \caption{The average distance between the nearest points from Place Pulse 2.0 data \citep{dubey2016deep} and our approach, according to their polarity in each city and neighborhood. [Best in color]}
    \label{fig:overlapping_distance}
\end{figure}

Since there is no direct mapping between our perception classes and the ones considered by Place Pulse, we aggregated them according to their sentiment polarity. This was done based on the description given by the authors of the project \citep{dubey2016deep}. Thus, we considered the Place Pulse categories wealthy, beautiful, and safety (because they represent places that look safer) as positive. Boring and depressing as negative. Moreover, the ``lively'' category as neutral, since it appears between positive and negative categories in our approach (see Figure \ref{uop-dict}). In this way, we performed a comparative analysis of perceptions extracted via our approach and Place Pulse, despite the different labels used in these approaches. 

First, we analyzed the spatial similarity between perception points from both approaches to understand how near they are. We do that as follows. Given a point $p$ with a particular polarity according to Place Pulse, we find among points with the same polarity extracted via our approach the one whose distance to $p$ is minimum. We do this process for all points from Place Pulse. According to this process, Figure \ref{fig:overlapping_distance} shows the average distance (with 95\% confidence interval) from all points, separated according to sentiment class and neighborhood for each city. As there are no points with perceptions in the Norwood Park neighborhood on Place Pulse data, we ignored this region in forthcoming results.

\begin{figure}[!htp]
  \centering
    \subfloat[Loop.]{
        \includegraphics[scale=0.28]{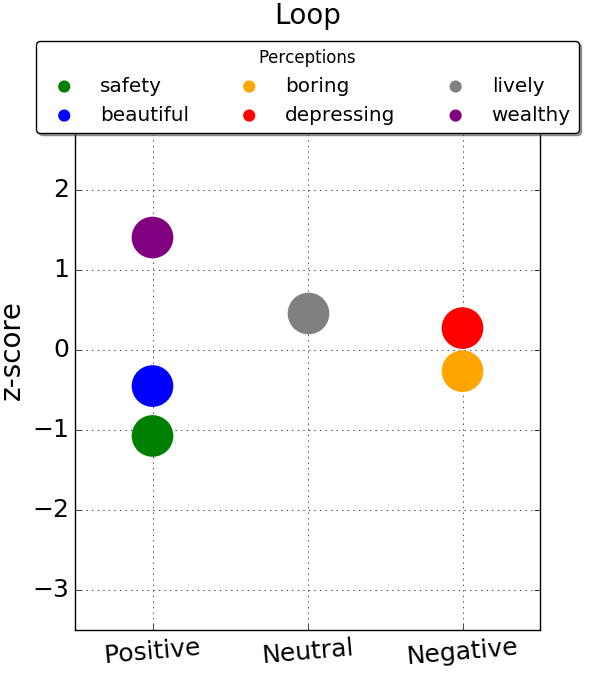}}
    \subfloat[Wicker Park.]{
        \includegraphics[scale=0.28]{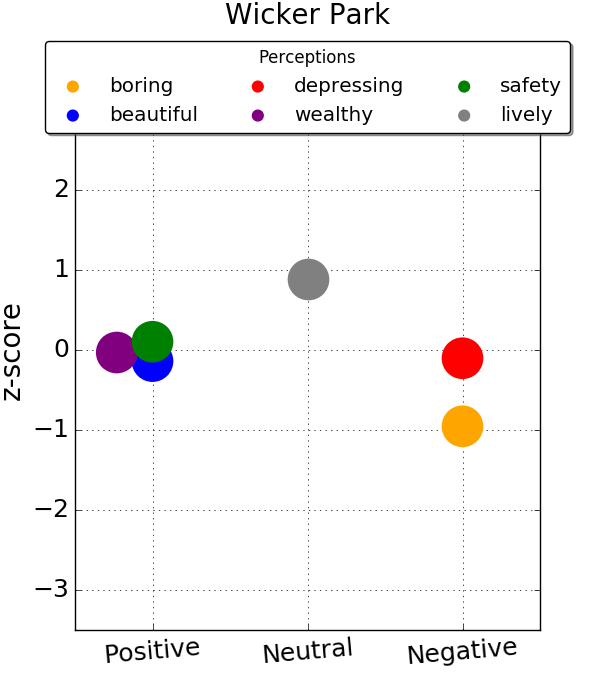}}
    \caption{The perception strength of Chicago's neighborhoods according to Place Pulse 2.0 data \citep{dubey2016deep}. [Best in color]}
    \label{fig:chicago-place-pulse}
\end{figure}

As we can see, the average distance follows a similar behavior in most cases. Positive images tend to have the shortest distances, followed by negative and neutral ones. One of the most prominent exceptions is Hammersmith and Fulham boroughs, where points with negative polarity have the most significant average distance compared to others. The average distance in all cases is around 100 to 200 meters for American cities. For London, the average distance range is higher, about 100 to 400 meters. This is comprehensible since the studied areas in London are considerably larger than the others. By this analysis, we observed that, in general, there are few spatial divergences between our approach and Place Pulse. This suggests that both approaches tend to agree reasonably well on the sentiment reflected by urban areas.

After that, we evaluated the perception strength using Place Pulse data based on $z$-score, as defined in Eq.\ \ref{eq:perc_strength}. We aggregated the perceptions according to their sentiment polarity to facilitate the comparison. Figure \ref{fig:chicago-place-pulse} shows the results for the evaluated areas of Chicago. According to Place Pulse data, Loop is mainly \emph{wealthy}, but is also \emph{lively} and \emph{depressing}, with less intensity as shown in Figure \ref{fig:chicago-place-pulse}(a). The similarity between the perception identified by our approach with this result is striking, where the most intense perception is also positive (\emph{SPECTACULAR}) followed by \emph{LIVELY} and negative perceptions. The perception about Wicker Park, Figure \ref{fig:chicago-place-pulse}(b), is predominantly \emph{lively}. Note that the positive perceptions are concentrated close to zero, being the safety perception the most significant among them. The negative perceptions also have less relevance in this region. According to our results, the \emph{LIVELY} category also stands out (third most prominent). At the same time, the \emph{SPECTACULAR} and \emph{RESPECTFUL} are the primary and secondary categories, respectively, with the negative perceptions keeping the average $z$-score below zero.

\begin{figure}[!htp]
  \centering
    \subfloat[Upper West Side.]{
        \includegraphics[width=0.3\textwidth]{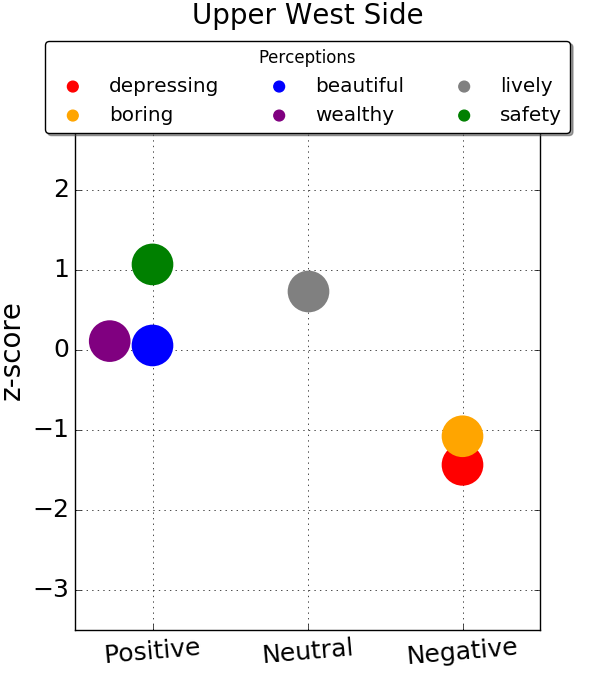}}
    \subfloat[Park Slope and Gowanus.]{
        \includegraphics[width=0.3\textwidth]{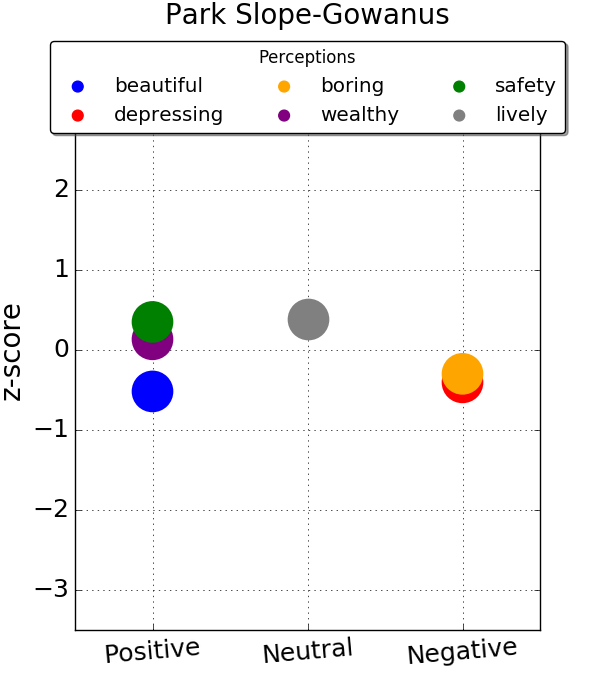}}
    \subfloat[Jamaica.]{
        \includegraphics[width=0.3\textwidth]{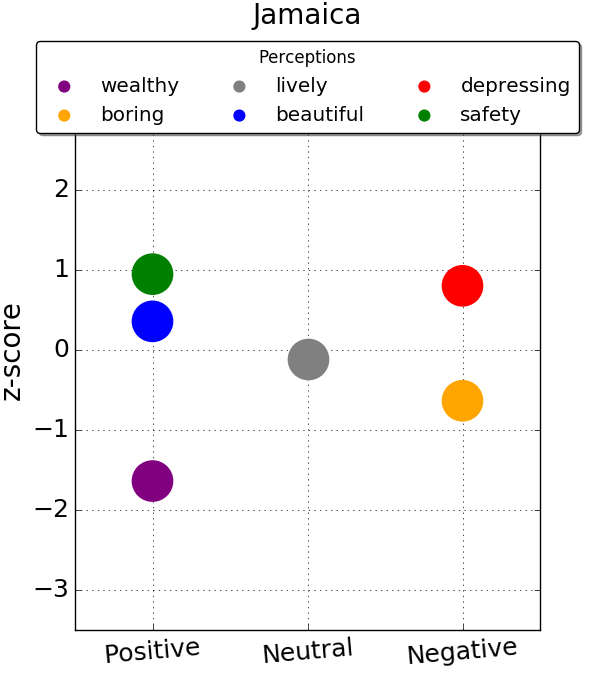}}
    \caption{The perception strength of NYC's neighborhoods according to Place Pulse 2.0 data \citep{dubey2016deep}. [Best in color]}
    \label{fig:ny-place-pulse}
\end{figure}

Similarly, Figure \ref{fig:ny-place-pulse} shows the results for evaluated areas of NYC. As shown in Figure \ref{fig:ny-place-pulse}(a), Upper West Side according to the Place Pulse data is mostly \emph{safety} and \emph{lively}, followed by \emph{wealthy} and \emph{beautiful}. Besides, the negative perceptions have low strength in this region ($z$-score $<$ -1). According to our results, positive perceptions also stand with strength close to zero, except category \emph{SPECTACULAR} that exceeds the others. Our results also identified low strength for negative perceptions in this area. Only the neutral category shows a slight difference, where \emph{LIVELY} is the fourth most evident perception for such region.  Nevertheless, the results show a strong correlation between the approaches. For Park Slope and Gowanus, Figure \ref{fig:ny-place-pulse}(b), all perceptions keep the $z$-score close to zero.  Thus, it is an area where no perception stands out. Note that our results have a significant match for the same area, following a similar behavior. The results for the Jamaica neighborhood, Figure \ref{fig:ny-place-pulse}(c), show a high intensity of \emph{depressing} perception, being the second most prominent perception in this area according to Place Pulse. Besides, \emph{safety} and \emph{beautiful} perceptions are the most relevant positive categories, while the \emph{lively} perception has $z$-score slight below zero. According to our results, a negative perception, \emph{AGGRESSIVE}, also has high strength in this area, followed by \emph{LIVELY} and \emph{GREAT}. Despite the divergence on the rank order of perceptions, the overall picture is still well captured.

\begin{figure}[!tp]
  \centering
    \subfloat[Hammersmith and Fulham.]{
        \includegraphics[width=0.3\textwidth]{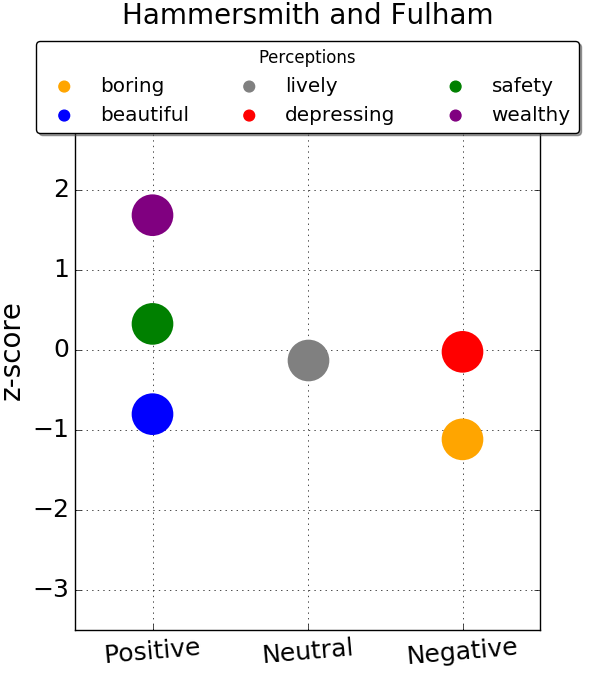}}
    \subfloat[Islington.]{
        \includegraphics[width=0.3\textwidth]{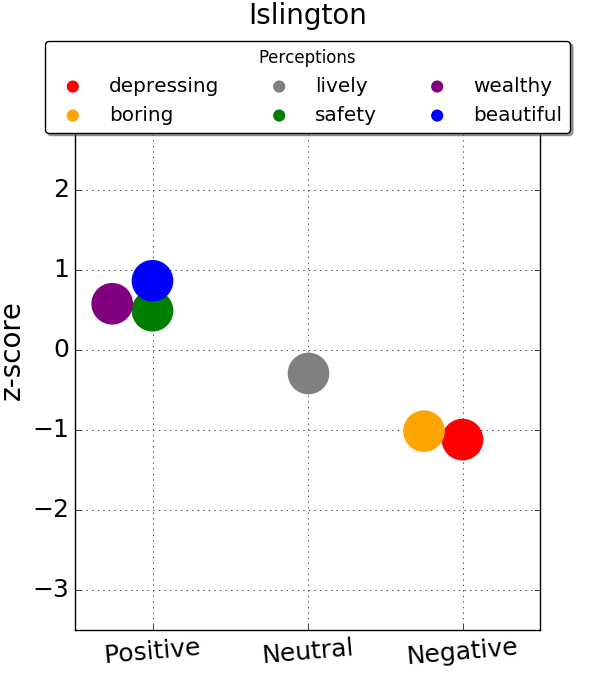}}
    \subfloat[Lambeth.]{
        \includegraphics[width=0.3\textwidth]{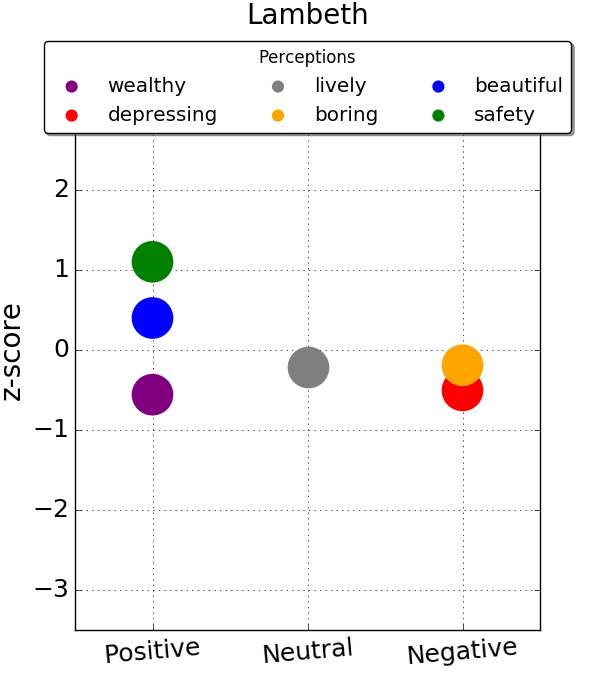}}
    \caption{The perception strength of London's neighborhoods according to Place Pulse 2.0 data \citep{dubey2016deep}. [Best in color]}
    \label{fig:london-place-pulse}
\end{figure}

Finally, Figure \ref{fig:london-place-pulse} shows the results for London. According to Place Pulse, the perception about Hammersmith and Fulham is predominantly \emph{wealthy} and \emph{safety}, followed by \emph{depressing} and \emph{lively}, as shown in Figure \ref{fig:london-place-pulse}(a). Our approach also identified a very similar perception for this area, where the \emph{SPECTACULAR} category corresponds to a primary perception of the neighborhood. At the same time, \emph{LIVELY} is a secondary category. Furthermore, our results show that the \emph{AGGRESSIVE} category has a similar strength to other positive perceptions. This indicates that it might coexist, as the case of the \emph{depressing} perception, according to Place Pulse in this area. For Islington borough, Figure \ref{fig:london-place-pulse}(b), the positive and neutral perceptions have a good match between our results and Place Pulse. However, according to our results, this area also suffers the influence of negative perceptions, mainly the \emph{WRONG} and \emph{CREEPY} categories. In contrast, Place Pulse results show low relevance of negative perceptions in the same region. Similarly, our results show a high intensity of several positive, neutral and negative perceptions at Lambeth.

In general, our approach, compared with the Pulse Place data, achieved a satisfactory similarity between extracted perception by both strategies in most cases.

\section{Discussions and Limitations}
\label{sec:discussion}
Our work has the potential to help people to extract knowledge from the city, and, thus, to improve its understanding. This is useful for many tasks, for instance, to assist in the development of intelligent services, such as personalized route recommendations, which could be offered by systems like Waze or Google Maps. In this context, tourists can explicitly request routes to walk through areas with selected perceptions.

It is also important to note that our approach could be applied, with the proper adaptations, to other domains as well. In this study, we focused on urban outdoor areas, but our approach could be explored, for example, for indoor spaces. By employing our methodology in other review datasets, perhaps larger and more updated than those explored, we could potentially enrich our dictionary with new significant words. This is important because the language is in constant evolution and can change in unique ways in different locations.

Place Pulse 2.0 considers visual patches of images to identify the place's urban perception instead of free-texts shared by individuals while visiting the area, which can potentially differ the people's perception since that non-visual features can influence in their perception. However, it is one of the few studies that mapped the urban perception in a large scale, considering a significant number of categories, and provided their data, giving us a piece of general knowledge about urban perception of outdoor areas.

We are also aware of some limitations of our proposal. Upon analyzing samples of messages shared by individuals, we found certain phrases containing conflicting perceptions. For example, ``\textit{... What an energetic, \textbf{amazing}, fun, \textbf{loud}, show ....}'' and ``\textit{... For anyone having a \textbf{bad} week or needs a \textbf{happy} break, here's a polar bear ...}''. In those examples, positive and negative perceptions would be identified. Our approach neutralized this problem with the proposed clustering step; however, if this type of situation happens many times in the same area, this could favor identifying wrongly two types of perceptions, depending on the values for the approach's parameters. We did not find any relevant problem regarding these cases in our results. However, the application of our approach in other datasets might face this possible limitation.

Besides, our approach takes into account the opinions of Twitter users who shared their perceptions when visiting the evaluated areas. Despite the clear advantages of using social media data to extract the overall picture of urban areas' perceptions, such findings might not necessarily correspond to accurate truth. This is because just a few population groups are more likely to use social media, especially adults (18+), urban dwellers, and high-income people \citep{pew2019}. Moreover, the human perception spectrum arising from the data sources considered in our study can be less expressive than those that could have been obtained by traditional surveys. However, note that other data sources, perhaps more expressive towards specific perceptions, could also be explored by our methodology, and thus, being complementary instead of replacing them.

\section{Conclusion}
\label{sec:conclusion}
In this study, we presented an automatic approach to support the learning and mapping of perceptions of urban outdoor areas from an extensive collection of noisy data expressing individuals' opinions in LBSNs. Due to several advantages of our approach, e.g., easier scalability, it has the potential to complement traditional methods, such as surveys. In this sense, scenarios where traditional surveys have been conducted, our approach can take advantage of their data (assuming availability) to learn about citizens' perceptions and proceed with our methodology's steps. Also, in scenarios where there is no information about urban perceptions, our approach can support bring such information, helping people better understand the semantics existing in different city regions. As future work, we intend to evolve this approach to incorporate other data sources, such as Instagram and Facebook. Besides, we plan to apply and evaluate our strategy for content in different languages and analyze more details about the implications of other variables, such as weather, on the results.

\begin{acknowledgements}
This work was partially supported by Coordena\c{c}\~{a}o de A\-per\-fei\-\c{c}o\-a\-men\-to de Pessoal de N\' {i}vel Superior -- Brasil (CAPES) -- Finance Code 001 (process 88881.337131/2019-01), project URBCOMP (Grant \#403260/2016-7 from National Council for Scientific and Technological Development agency - CNPq), and GoodWeb, BigCloud and Mobilis (Grants \#2018/23011-1, \#2015/24494-8 \& \#2018/23064-8 from S\~{a}o Paulo Research Foundation - FAPESP). 

\end{acknowledgements}



\bibliographystyle{spbasic}
%\bibliography{mybib}

%------INSERINDO A BIBLIOGRAFIA SEM O ARQUIVO BIB

%---------------

\clearpage
\appendix
\section{Evaluated Areas} \label{apped_a}

\begin{figure}[!htp]
   \centering
    \includegraphics[scale=0.45]{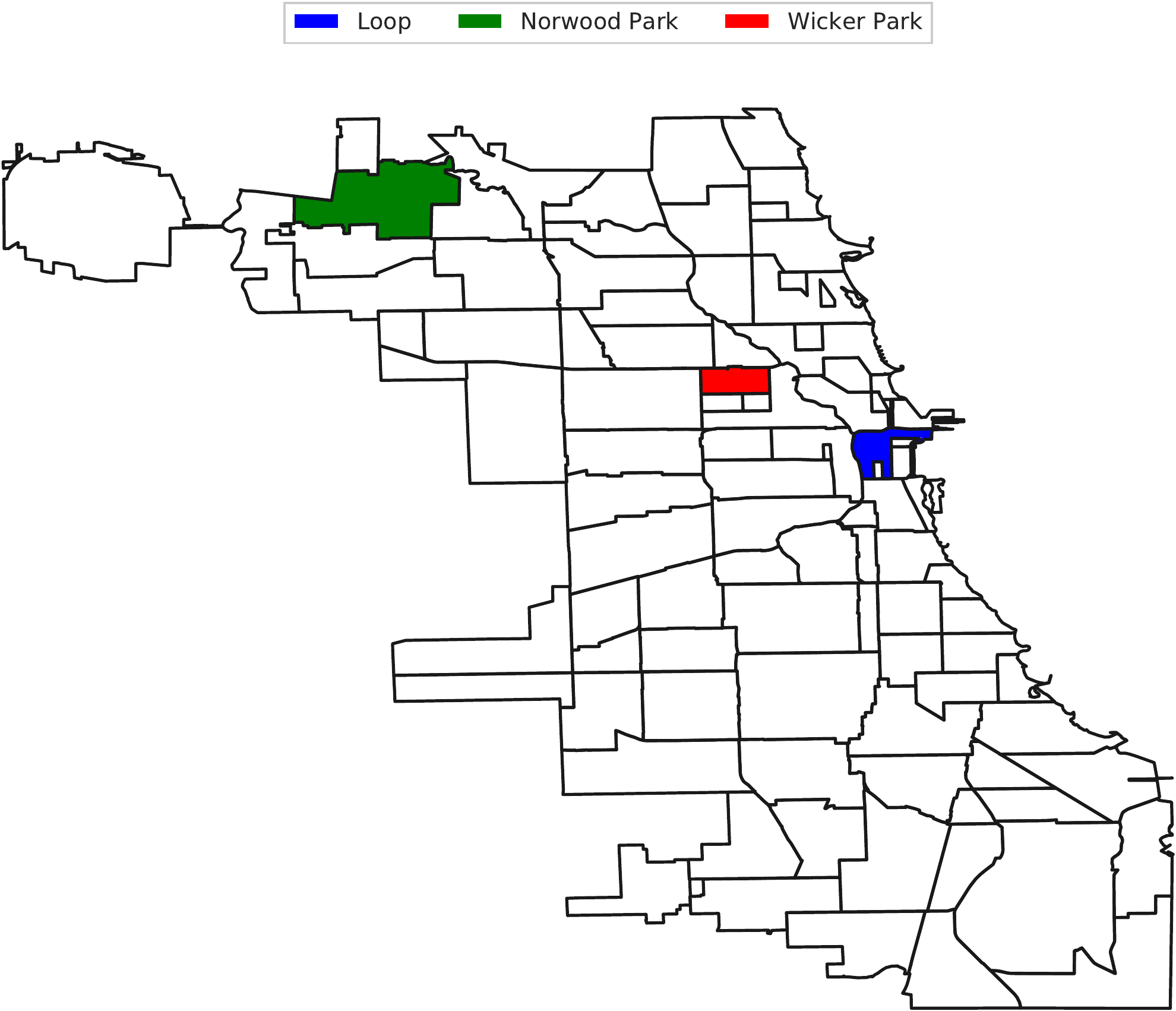}
    \caption{Evaluated neighborhoods in Chicago, IL.}
    \label{fig:eval-scenarios-chicago}
\end{figure}
\begin{figure}[!htp]
   \centering
     \includegraphics[scale=0.45]{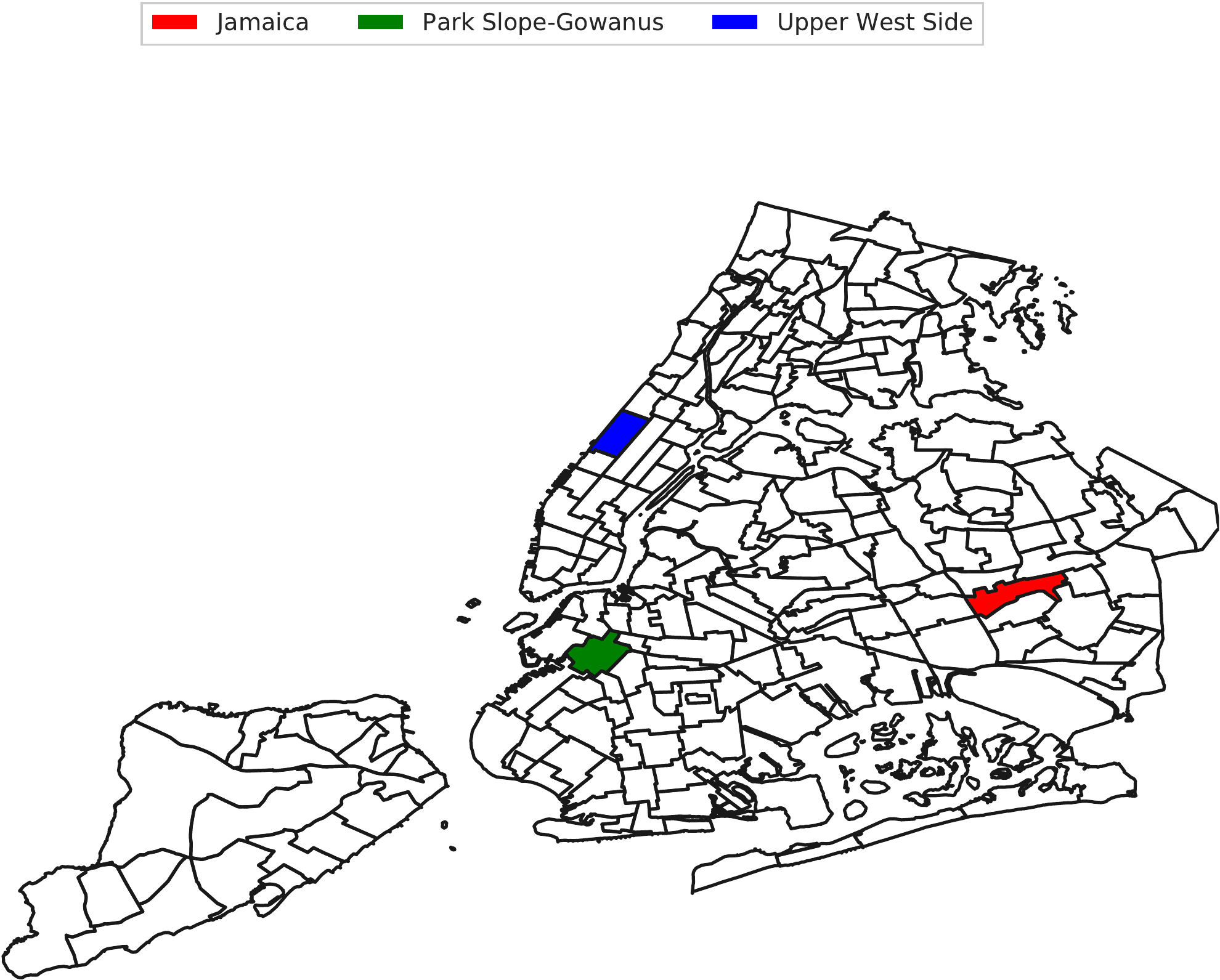}
     \caption{Evaluated neighborhoods in NYC, NY.}
   \label{fig:eval-scenarios-nyc}
\end{figure}
\begin{figure}[!htp]
   \centering
     \includegraphics[scale=0.45]{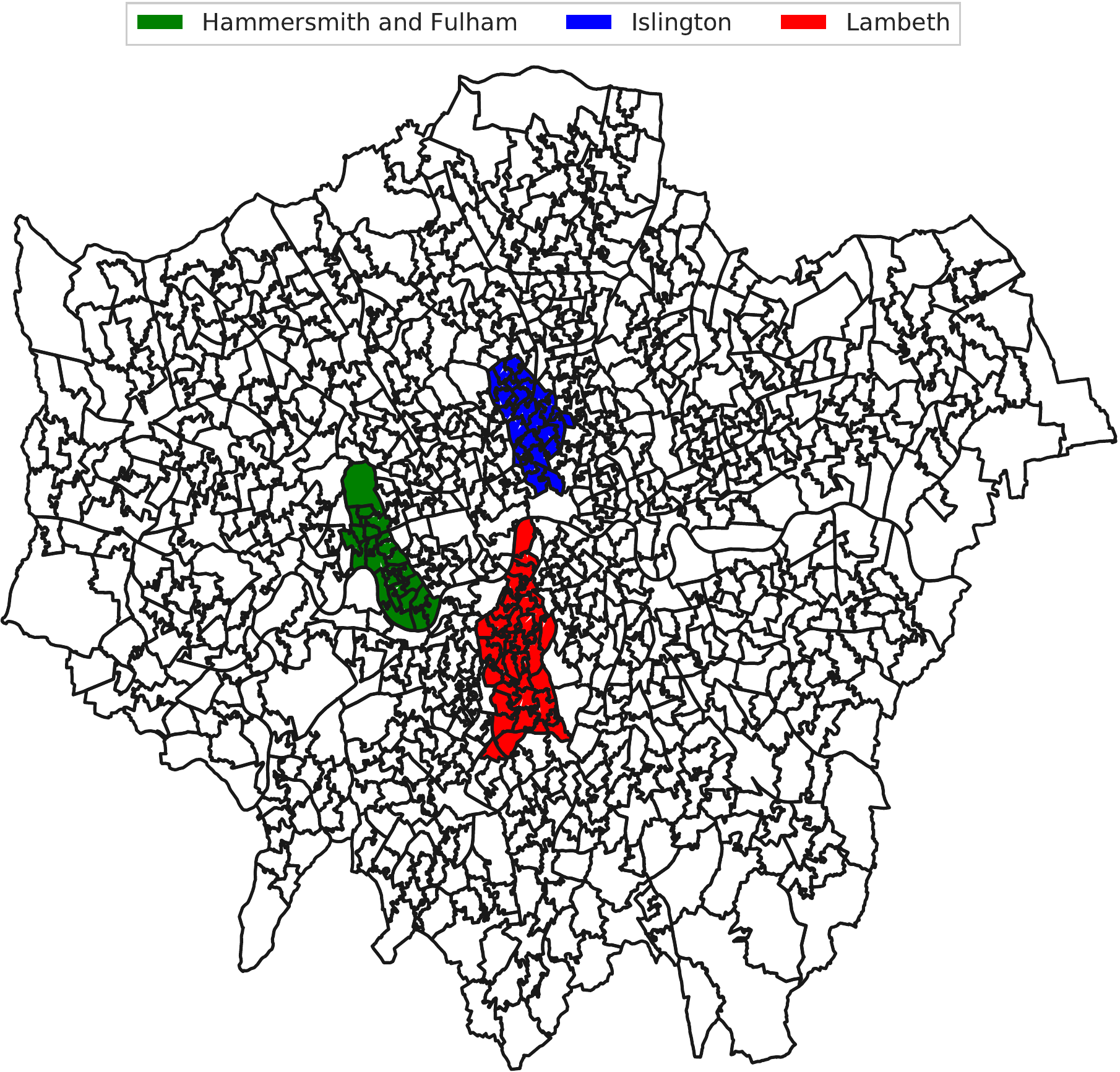}
     \caption{Evaluated neighborhoods in London, UK.}
   \label{fig:eval-scenarios-london}
\end{figure}
\clearpage
\section{Urban Perception Maps} \label{apped_b}

\begin{figure}[!htp]
   \centering
    \includegraphics[scale=0.45]{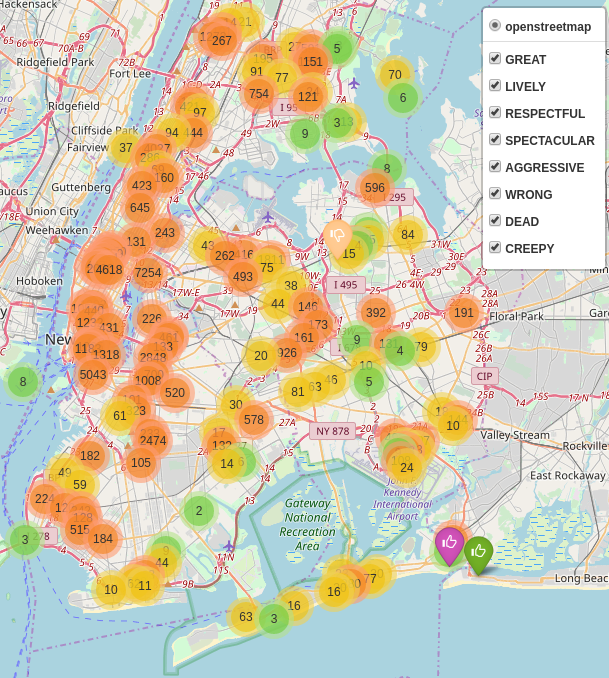}
    \caption{Screenshot of an interactive urban perception map in New York City. .}
    \label{fig:perception-maps-nyc}
\end{figure}
\begin{figure}[!htp]
    \centering
    \includegraphics[scale=0.45]{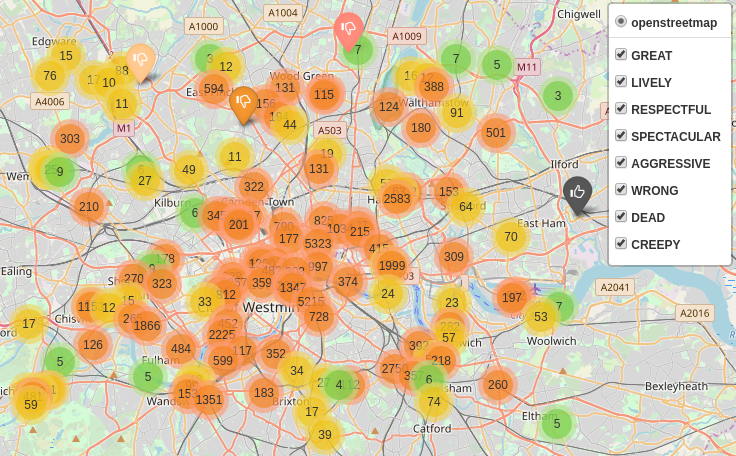}
    \caption{Screenshot of an interactive urban perception map in London. }
    \label{fig:perception-maps-london}
\end{figure}

\end{document}